# Defining Thixotropy

Yogesh M. Joshi[1] and Ronald G. Larson[2]

[1] Department of Chemical Engineering, Indian Institute of Technology Kanpur, Kanpur, Uttar Pradesh 208016, India. joshi@iitk.ac.in

[2] Department of Chemical Engineering, University of Michigan, Ann Arbor, MI 48109 rlarson@umich.edu

**Abstract**

Originally coined 100 years ago to represent a shear-reversible sol-gel transition, the range of the term "thixotropy" has gradually expanded to encompass time-dependent changes in viscosity. Since this risks confusing "thixotropy" with ordinary nonlinear viscoelastic behavior, in this work we seek simple, historically faithful, quantifiable, and experimentally testable definitions of "thixotropy" that avoid this confusion. By analysing both historical tradition and published experimental data, we arrive at three complementary definitions. Definition 0 provides an idealized, purely inelastic reference useful primarily for recognizing thixotropy in a constitutive equation. Definitions 1 and 2, which are more practical, require, respectively, unusually large strains to reach steady state, and unbounded growth of the material relaxation time during rest. Definition 1 can apply in either shear-rate or stress-controlled modes, with the latter encompassing the so-called "viscosity bifurcation." These criteria are tested across diverse materials that we identify as thixotropic based on their responses clearly exceeding their viscoelastic reference scales. The framework is deliberately phenomenological and does not require specific microscopic mechanisms, allowing diverse materials to be described within a common rheological language, as is typical for rheological terminology. We believe that these definitions provide a practical framework for assigning thixotropy a more definite place in rheological taxonomy.

## I. Introduction

"Thixotropy" is understood to be a rheological phenomenon in which a material's resistance to deformation and flow depends on time and flow history [1, 2]. The term was first introduced nearly a century ago by Pétérfi [3] to describe a spontaneous sol – gel transition of a suspension that can be reversed by mechanical agitation. The term "thixotropy" was derived from the Greek words "thixis," meaning stirring or shaking, and "trepo," meaning turning or changing, suggesting a change in the state of a material from gel-like to liquid-like upon shaking [1]. This rheological behavior is a consequence of a microstructure that progressively breaks down under sustained deformation, causing viscosity to decrease. Upon cessation or reduction of the deformation rate, the microstructure rebuilds, causing viscosity to recover [2, 4]. The original terminology proposed by Pétérfi thus captured the essence of materials that undergo reversible structural transformations under mechanical stimulation, with accompanying rheological changes. Thixotropic materials are ubiquitous in industrial applications, and include paints and coatings, drilling muds, cosmetics and pharmaceuticals, food products, and construction materials [1, 2]. From a practical point of view, thixotropic behavior governs processing, handling, and end-use performance of the materials.

However, around the same time that the category of "thixotropic" materials was identified, polymeric materials were recognized as possessing "viscoelasticity," arising from a combination of rubbery-like entropic elasticity at short times and liquid-like flow at longer times [5]. The rapid growth in rheological measurement of polymeric materials, and the accompanying development of theories of polymer viscoelasticity by pioneers such as Flory, Kuhn, Rouse, Zimm, Ferry, Doi, and Edwards, and others [6], established the concept of viscoelasticity without a clear distinction from "thixotropy." As long as viscoelasticity pertained to polymers and thixotropy to suspensions exhibiting sol-gel transitions there was little need to clearly distinguish the two concepts. However, the study of more complex materials under the general framework of "soft matter" or "complex fluids," including polymeric glasses [7], waxy crude oils [8-10], emulsions, paints, gels, drilling fluids, and many others, has led to confusion, and to attempts to identify more clearly "thixotropy" as a rheological phenomenon distinct from "viscoelasticity" for arbitrary materials. A table of these and many other materials labeled “thixotropic” in the literature can be found in Larson and Wei [11]. A seminal contribution to the field is the work of Metzner and coworkers [12] who not only tabulated many of the thixotropic structural models, but integrated thixotropic effects with plastic and viscoelastic behavior in a model that they applied successful to a particle-filled melt.

The problem of rheological taxonomy is fraught, since both “thixotropy” and “viscoelasticity” involve time-dependent rheological phenomena. Thus, after a century of study, the concept of thixotropy lies at a crossroads, lacking an agreed-upon definition that discriminates it clearly from viscoelasticity [11]. Many recent papers have addressed this issue, but convergence on a definition has not so far been forthcoming [11, 13-17]. The purpose of this manuscript is not to try to resolve this knotty problem, but to outline alternative definitions, and their advantages and disadvantages, that researchers and practitioners can consider. We hope that, even if a single definition cannot at present be agreed upon, at least a set of clear alternatives can be identified, and a method of arriving at, or improving, definitions can be illustrated. At the minimum, we hope this article prompts practitioners to more clearly identify their definition when using the term “thixotropy.”

*Brief History:* Schalek and Szegvari [18] were the first to report that an iron-oxide sol transforms into a coagulated gel when at rest and becomes liquid again upon shaking, a process that could be repeated. Péterfi [3] termed this phenomenon of reversible microstructural rearrangements as “thixotropy.” The subsequent early papers on thixotropy, including those by Freundlich and Rawitzer [19], Hauser [20], Freundlich [21], and Goodeve and Whitfield [22], anchored thixotropy firmly to a reversible sol–gel transition: certain colloidal gels becoming fluid when agitated and reforming into a gel when left undisturbed. This suggests that the original conception was unambiguously structural and was phenomenologically distinct from the notion of viscoelasticity. However, the middle decades of the twentieth century led to the widening of the scope of the concept of thixotropy. During that period, when rheological taxonomy was still evolving, some authors treated thixotropy merely as a time-dependent behavior and, in doing so, placed emphasis on temporal changes in a macroscopically measurable parameter such as viscosity. Consequently, a decrease in viscosity with time under shear and an increase in the same when at rest became a measurable manifestation of "thixotropy" rather than an explicit reversible sol–gel transition. In addition, expanding industrial interest in materials showing thixotropic behavior needed a simple diagnostic test, which, in the absence of rigorous discrimination protocols, gradually led to the adoption of operational definitions based on observing time-dependent viscosity during and after an applied deformation field. Such "thixotropic" behavior no longer required a reversible gel–sol transition and no longer insisted on a clear microstructural mechanism. Over the years, this operational view became ingrained in the academic discourse, leading to the dilution of the original concept [1].

Now, shear thinning, which is a reduction of shear viscosity upon increase in shear rate, is a common phenomenon, and for polymers is conventionally attributed to nonlinear viscoelasticity [23, 24]. However, shear thinning in polymers takes a finite time to fully manifest after the start of shearing and is gradually reversed over a finite time when the deformation field is removed [13, 25]. Since a "viscosity decreasing with time under shear" is phenomenologically similar whether it occurs in a polymer solution or in a suspension undergoing "microstructural breakdown toward a sol-like state," it is evident that the conceptual barrier between thixotropy and viscoelastic shear-thinning has weakened over time. The consequence of this is that the original concept of thixotropy has been partially overshadowed by broader, less discriminating definitions whose scope overlaps with standard viscoelastic transients. This historical drift is a major reason why modern literature contains substantial inconsistency in the use of the term "thixotropy" [11], and why a return to a more principled, historically aware definition is necessary.

**Table 1**: Chronological evolution of definitions of Thixotropy (in verbatim, wherever available)†

| | |
|---|---|
| T. Péterfi (1927) [3] | Thixotropy is the phenomenon that a coagulated sol can be liquefied by merely shaking and always again coagulated to a gel by stopping agitation. |
| H. Freundlich and W. Rawitzer (1927) [19] | By thixotropy is meant the phenomena of concentrated gels which solidify to gels which may again be liquified to sols. |
| | The resolidification occurs repeatedly, at constant temperature with a constant speed. |
| J. Pryce-Jones (1936) [26] | an increase of viscosity in a state of rest and a decrease of viscosity when submitted to a constant shearing stress. |
| Scott Blair (1944) [27] | With our present knowledge of microstructural changes, it is probably safe to say that all materials that are shear thinning are thixotropic, in that they will always take a finite time to bring about the rearrangements needed in the microstructural elements that result in shear thinning. If this recovery is very rapid, the phenomenon is observed as shear thinning; if slow, it is observed as thixotropy. |
| Carman (1949) [28] | Reversible gelation is known as thixotropy |
| Oxford Encyclopedia Dictionary of Physics (1962) [29] | Certain materials behave as solids under very small, applied stresses but under greater stresses become liquids. When the stresses are removed the material settles back into its original consistency. This property is particularly associated with certain colloids which form gels when left to stand but which become sols when stirred or shaken, due to a redistribution of the solid phase. |
| Bauer and Collins (1969) [30] | When a reduction in magnitude of rheological properties of a system, such as elastic modulus, yield stress, and viscosity, for example, occurs reversibly and isothermally with a distinct time dependence on application of shear strain, the system is described as thixotropic. |
| Reiner and Scott-Blair [31] | Property of a body by virtue of which the ratio of shear stress to rate of deformation (viscosity) is temporarily reduced by previous deformation. |
| Morris (1969) [32] | The property exhibited by certain gels of liquefying when stirred or shaken and returning to hardened state upon standing. (There are several definitions that are very similar |
| | to this one in various dictionaries, differing only by minor paraphrasing.) |
| Chambers dictionary of science and technology (1971) [33] | Rheological property of fluids and plastic solids characterized by a high viscosity at low stress, but a decreased viscosity when an increased stress is applied. |
| Van Nostrand's scientific encyclopedia (1976) [34] | A thixotropic fluid is a fluid whose viscosity is a function not only of the shearing stress, but also of the previous history of motion within the fluid. The viscosity usually decreases with the length of time the fluid has been in motion. |
| IUPAC (1979) [35] | The application of a finite shear to a system after a long rest may result in a decrease of the viscosity or the consistency. If the decrease persists when the shear is discontinued, this behavior is called work softening (or shear breakdown), whereas if the original viscosity or consistency is recovered this behavior is called thixotropy. |
| Barnes, Hutton and Walters (1989) [36] | Decrease (in time) of viscosity under constant shear stress or shear rate, followed by a gradual recovery when the stress or shear rate is removed. |
| Birley (1991) [37] | Time-dependent fluid behavior in which the apparent viscosity decreases with the time of shearing and in which the viscosity recovers to, or close to, its original value when shearing ceases. The recovery may take place over a considerable time. |
| Chambers 20th Century Dictionary (1993) [38] | The property of gels of showing a temporary reduction in viscosity when shaken or stirred. |
| Polymer Technology Dictionary (1994) [39] | Thixotropy. A term used in rheology which means that the viscosity of a material decreases significantly with the time of shearing and then, increases significantly when the force inducing the flow is removed. |
| IUPAC, Compendium of Chemical Terminology: The Gold Book (1997) [40] | The application of a finite shear to a system after a long rest may result in a decrease of the viscosity or the consistency. If the decrease persists when the shear is discontinued, this behavior is called work softening (or shear breakdown), whereas if the original viscosity or consistency is recovered this behavior is called thixotropy. |
| Coussot, Bonn and coworkers (2002) [41] | Thixotropic yield stress materials exhibit a viscosity bifurcation behavior, wherein there is a critical shear stress below which the material will eventually stop flowing, while above which it leads to steady-state flow. This critical stress is identifiable as a time-dependent yield stress. The shear rate at this critical stress is finite and corresponds to the lowest shear rate for which homogeneous flow can be observed. |
| Mewis and Wagner (2009. 2012) [2, 4] | While Mewis and Wagner adopt the IUPAC definition of thixotropy, they clearly state that nonlinear viscoelastic effects such as shear thinning should be excluded from its scope. They also include reversible physical aging and mechanical rejuvenation observed in soft glassy materials within the purview of thixotropy. |
| Larson (2015) [25] | Ideal thixotropy refers to a purely viscous, time-dependent response, wherein viscosity increases with rest time and decreases under applied deformation field. In ideal thixotropic response, the stress depends on the history of the strain rate but there is no contribution from elasticity.<br><br>A non-ideal thixotropic fluid retains all the features of ideal thixotropy but also exhibits some viscoelasticity, provided that the viscoelastic relaxation is much faster than the thixotropic evolution. |
| Joshi and co-workers (2021) [13] | An increase in relaxation time with waiting time since cessation of preshear, provided the material also shows shear rejuvenation, serves as a defining criterion for thixotropy. |

The drift in understanding of the term "thixotropy" is made evident in Table 1, a chronological list of definitions and descriptions of "thixotropy" starting from 1927 and continuing to the present. Despite their variety, the various definitions mentioned in the table can be grouped mainly into two classes. The first class, which was prevalent for the first two or three decades since the term was introduced, describes thixotropy as a shear-reversible sol–gel transition. The second class, which became ever more dominant since the 1970s, invokes viscosity and its corresponding increase or decrease under both quiescent and deforming conditions. The issue with definitions in the first class is that, unlike "shear-thinning," "yield stress," or even "viscoelasticity," they do not specify a rheological test that clearly defines the presence of thixotropy. The second class of definitions, invoking time-dependence in the shear thinning and subsequent recovery of viscosity, risks overlapping with nonlinear viscoelasticity. A few notable exceptions include the definition by Bauer and Collins [30], which explicitly mentions time-dependent changes in elastic modulus and yield stress, in addition to viscosity, something no earlier definition had required. Larson [25] introduced the concept of ideal thixotropy to clearly separate it from the domain of nonlinear viscoelasticity.

Based on how one defines thixotropy, the literature recommends various kinds of rheological tests to ascertain its presence. The main theme of these is that, by using rheological measurements, thixotropy is recognized by the destruction of a microstructure under an applied deformation field, and the rebuilding of the same when the deformation field is removed. One of the popular tests is the presence of a hysteresis loop when the shear rate is first increased (or decreased) from a low (or high) value and then reversed at the same rate. The area enclosed by the loop is often taken as a measure of thixotropy [42-45]. However, the nature of the loop, clockwise or anticlockwise, and the loop area depend on the applied shear-rate limits and the ramp rate. While the hysteresis loops are very popular in identifying the thixotropic nature of a material, the limitation of this method is that the hysteresis loops have also been observed in many viscoelastic materials and hence cannot be considered as a definition or a definitive test for thixotropy [14, 46, 47]. Another test used to verify the presence of thixotropy is the shear-rate step-down test [2, 4]. If, in response to this test, the stress exhibits an undershoot (a decrease followed by an increase) before reaching the steady value associated with the final shear rate, the material is deemed thixotropic. However, nonlinear viscoelastic materials also show such undershoots [13], indicating that this behavior cannot be attributed solely to thixotropy. A related complication arises when the stress instead first increases and then decreases at a fixed shear rate following a step-down; Wang and Ewoldt [48] showed, using orthogonal superposition rheometry on a carbon black suspension, that such a signature can reflect genuine antithixotropic structuring rather than a purely thixotropic or viscoelastic response. Other tests, such as an increase in magnitude of a stress overshoot in a shear startup experiment as a function of waiting time, are also found in conventional viscoelastic materials, such as polymer melts and solutions [13]. The practical issue associated with many diagnostics tests is that they conflate viscoelastic kinetics with genuine "structural kinetics," unless tests specifically probe rebuild at no flow (or at controlled, weak flow) over systematically varied waiting times with controlled and repeatable initial state. We believe that this experimental degeneracy is the heart of the definitional problem that this manuscript addresses.

Notwithstanding these overlaps, some behaviors remain more naturally associated with the classical notion of thixotropy and are uncommon in ordinary nonlinear viscoelasticity. The original idea of thixotropy, as understood by the founders and early researchers, points to a mechanically induced, reversible sol–gel transition under isothermal conditions. This notion can be divided into two parts. The first part concerns spontaneous gel formation under quiescent conditions (no flow and no stress). If a gel is viewed as a space-spanning network with inherently infinite viscosity, then its formation implies a time-dependent evolution of viscosity that diverges toward infinity. Such divergence also suggests that the material may eventually develop a yield stress. Moreover, because the network carries a finite elastic modulus, continued structural buildup may naturally lead to a time-dependent increase in both the elastic modulus and the yield stress. The term "yield stress" itself carries considerable ambiguity between its rheological and solid-mechanics usages, an ambiguity sharpened further in thixotropic materials showing viscosity bifurcation [49]. All three growths, namely the eventual divergence of viscosity, which could be followed by increases in elastic modulus and yield stress, may be regarded as classical signatures of thixotropy that do not typically appear in conventional nonlinear viscoelastic behavior. The second part of the original notion is that this buildup of viscosity and yield stress is at least partially reversible upon application of mechanical deformation. This requirement excludes chemical curing from the scope of thixotropy. Taken together, these two aspects may guide a definition of thixotropy that aligns closely with its classical understanding. The limiting case of ideal thixotropy, which fully excludes viscoelastic effects, also remains an important candidate in this discussion.

Beyond the need to respect the historical definitions and usages of the term "thixotropy," there are some other principles usually employed in rheological taxonomy. To clarify, one can consider how the rheological category of "shear thinning" satisfies all of these.

1. Simple – The rheological property is described in ideal form, so that each property is distinguished from other properties that might accompany it in a real material. Thus, the same material may possess multiple rheological properties.
2. Qualitative, but quantifiable – The existence of the property is distinguished by type, not by degree, but the degree to which the property exists can be quantified, for example by power-law index, relaxation time or modulus.
3. Mechanical and phenomenological – The categories are defined entirely by their continuum rheological response and not by material composition or structure.
4. Practical – while all materials, including water, are ultimately viscoelastic if frequency or time is infinitely resolvable, rheological properties are chosen for their usefulness.

Given these general principles and the example of shear-thinning, which meets them all, one can immediately see the appeal of defining thixotropy as "time-dependent, reversible, shear thinning." This closely matches several of the historical definitions, and could be one option, as long as one is prepared to include many forms of nonlinear polymer viscoelasticity as "thixotropy." Another alternative is to restrict "thixotropy" to particular material types, such as suspensions, thus violating principle 3, but holding more closely to the historical origins of the term "thixotropy." Other alternatives involve various levels of complexity, requiring, besides "time-dependent shear thinning," other characteristics, or the absence thereof, to deem a material thixotropic. Adding additional specifications risks violating Principles 1, 2, or 3, however.

Building on the above considerations, and the historical and mechanistic perspectives, we here propose a practical, tiered framework of three definitions two of which can be directly applied to experimental measurements. In Section II, we present the definitions, note the associated caveats, and discuss the underlying reasoning. Section III shows example experimental results from the literature that meet Definitions 1 and 2. Finally, Section IV contains concluding remarks. Our aim is to offer a pragmatic vocabulary for researchers and industry that respects the historical evolution of the concept of thixotropy while remaining robust to the experimental and modelling ambiguities that have accumulated in the literature.

## II. Proposed definitions:

As emphasized in the Introduction, the term "thixotropy" has gradually drifted over the decades from its original proposal of flow reversible sol–gel transitions to a more general version that includes time dependence of viscosity under quiescent as well as flow conditions. This generalization, if not further clarified, is more ambiguous than the earliest definitions, as it can end up including nonlinear viscoelastic transients and simple shear-thinning behavior within "thixotropy".

To address this issue, and to cover the different types of rheological responses that are deemed to display thixotropy, we suggest several definitions of "thixotropy," following guiding considerations drawn from the above discussion and the four general principles of rheological taxonomy listed above. First, the definitions should be faithful to the historical context. Second, the definitions must be relatively simple, and yet able to distinguish thixotropy from nonlinear viscoelasticity. Third, the definitions must be expressible quantitatively in continuum rheological terms such as stress, strain rate, viscosity, relaxation time, elastic modulus, yield stress, etc. without requiring any material-specific microscopic details. Fourth, the definitions should be practical, in the sense that they should lead to measurable and reproducible criteria for identifying thixotropic behavior from rheological experiments.

With these guiding principles in mind, we now propose a set of three alternative definitions. These definitions are not intended to be mutually exclusive. Instead, they emphasize different limiting perspectives that naturally arise in experimental practices. Together, they outline the breadth encompassed by the concept of thixotropy over past century while clarifying its distinction from nonlinear viscoelasticity.

Definition 0: *Thixotropy refers to a gradual decrease in viscosity during the start-up of shear following a prolonged period of rest or weak flow, and a subsequent increase in viscosity after cessation or step-down of shear rate, with negligible contribution from elasticity.*

This definition reflects an idealized thixo-viscous or approximately inelastic thixotropic limit, in which time dependence arises from reversible structural evolution affecting viscous resistance, with little contribution from elastic energy storage. The stress, therefore, relaxes quickly upon removal of flow. Definition 0 represents a clean conceptual baseline, closely aligned with classical "ideal thixotropy" discussed in the literature [25]. *It is most clearly applied to rheological constitutive equations, for which "thixotropy" is recognized by the completely instantaneous relaxation of stress on flow cessation, and the complete absence of any elastic recoil when stress is removed.* Thus, regardless of any disputes over what qualifies as a thixotropic material in the laboratory, “pure” or ideal thixotropy is a conceptual reality. Constitutive equations can be, and are, written that express this ideal. Just as a simple “Bingham fluid” [50] represents a useful approximation for solving idealized problems, “ideal thixotropy” can likewise be a useful approximation. Although ideal thixotropic materials do not exist in reality, they do exist in the human mind and in constitutive models, and this is enough reason to define the concept of thixotropy, however imperfectly it may be realized in actual materials.

For real materials, the primary strength of this definition lies in its discriminatory clarity. By explicitly requiring “negligible” elasticity, Definition 0 excludes nonlinear viscoelastic responses that could otherwise be classed as thixotropy in common rheological tests. However, it is overly restrictive since many materials that are conventionally considered thixotropic do exhibit a significant linear and nonlinear viscoelastic response, especially at a matured state of microstructure build-up. Consequently, Definition 0 should be viewed primarily as a conceptual and modeling ideal, rather than a universal descriptor of thixotropy in real experimental materials. By itself, it cannot be used to discriminate the full range of thixotropic phenomenology observed in practice. The responses of the real materials that can usefully be termed as thixotropic are often governed by a competition between "structure" formation and deformation-induced "structure" break-down, resulting in the possibility of arrested or flowing states, depending on the magnitude and nature of the deformation field. Here the term "structure" needs to be distinguished from polymer deformation, as represented for example by the "conformation tensor" **S**, used in nonlinear viscoelastic theories. Such conformation variables represent alignment of long-chain molecules, associated with highly elastic recoverable strain, and properly identified with nonlinear viscoelasticity rather than "thixotropy" [51, 52]. Since rheological properties are assigned based on phenomenology, rather than the microstructure, the challenge is to define thixotropy without reference to whether the material is polymeric, or colloidal, or some mixture of the two.

Many rheological protocols also impose deformation through a prescribed shear rate, stress, or strain history, including step jumps in imposed shear rate or applied stress. In such cases, the material response is more naturally interpreted in terms of the evolution of the conjugate variable, stress under rate control, or strain under stress control, as the material approaches, or fails to approach, a steady state. This necessitates a complementary description given here as Definition 1.

Definition 1: *Thixotropy is exhibited when, following an order of magnitude increase in deformation rate or stress, the strain required for attainment of steady state in a rheological property such as viscosity, relaxation time, and/or yield stress, is much larger than what can be attributed to nonlinear viscoelasticity.*

Definition 1 poses a single, clear question: when the deformation rate, or the applied stress, of a material is increased significantly, how much strain is required for the relevant rheological property to reach its new steady value? If this required strain is much larger than what a nonlinear viscoelastic material would typically need, the material is considered thixotropic according to Definition 1. For definiteness, "much larger" can be taken to mean "at least ten times larger." Note that here, strain, rather than time, is the key variable, since it's the accumulated deformation that causes microstructural breakdown, not simply the passage of time. This aligns with the classical view, noted under Definition 0, that viscoelastic responses reach the steady state at strains of roughly order unity (although we will see that long-chain polymers can require strains of 10 or even higher), while thixotropic restructuring requires prolonged deformation spanning from tens to thousands of strain units. Subjecting an initially undisturbed material to a step-up in shear rate is the simplest and clearest example of such deformation, especially since the initial shear rate is effectively zero.

In a step-shear rate jump, as suggested by Definition 1, it is necessary to stipulate that thixotropy should *not* be inferred when the long transients arise from inhomogeneous deformations. Some thixotropic materials show a non-monotonic steady-state flow curve, in which the steady-state shear stress decreases with increasing shear rate, typically in a limit of low shear rates. In this domain, a spatially homogeneous flow is not stable, and the material instead separates into coexisting bands, a phenomenon known as shear banding (discussed further below, in connection with the stress-controlled, bifurcating form of this same definition), with the relative widths of the bands set by the lever rule. Should either the lower or the higher shear rate associated with the step shear rate jump happen to fall within such a non-monotonic region of the flow curve, the corresponding "steady state" would not be a homogeneous material response, but rather some composite of coexisting bands. The strain needed to arrive at this composite state would then reflect band-width readjustment and interface migration, and not only the intrinsic structural kinetics that Definition 1 is intended to probe. It is, therefore, necessary that neither the lower nor the higher shear rate associated with the step lie in a region where the flow curve is non-monotonic. Both the shear rates should instead be chosen from regions where the steady state is monotonic, that is, where shear stress increases with shear rate. This ensures that the flow is homogeneous at both ends of the imposed step, and that the strain required to go from one homogeneous steady state to the other can be attributed, without ambiguity, to genuine thixotropic restructuring rather than to the migration of shear bands.

Definition 1 focuses solely on an increase in deformation rate or applied stress, and does not include recovery during rest or a step-down in either variable, for the following reasons. First, in a weaker but nonzero field, the shear rate or stress influences both the rate at which the microstructure rebuilds as well as the state of the final microstructure that is achieved, since the steady state is determined by the applied rate or stress. As a result, neither elapsed time nor accumulated strain provides a clear, protocol-independent measure of recovery. However, we would like to make it clear that Definition 1 does not exclude the possibility that a step-down in rate or a period of rest can also indicate thixotropic behavior; such observations are indeed valuable but are best treated as corroborating rather than as defining evidence for Definition 1 for the reasons mentioned above.

A further practical question is what should be regarded as the strain "attributable to nonlinear viscoelasticity," since this reference is not universal but material-specific. Where a molecular theory is available, it may supply this reference directly. For example, in the case of a well-entangled polymeric liquid, entangled polymer theories that include “chain stretching” [51] predict that strains of around 2 up to greater than 10 strain units (scaling as the Weissenberg number to a power of 1/3), are needed to reach a steady state during shear start-up, and this prediction is in good agreement with experiments [53]. A strain of order ten or even higher for such a system is therefore not by itself evidence of thixotropy, since it remains within reach of viscoelasticity for long-chain polymers. A strain approaching fifty or more, for entangled polymers, would be needed before viscoelasticity alone becomes implausible. For particulate and colloidal systems, which make up most materials conventionally regarded as thixotropic, no such theory is generally available. Therefore, the reference strain is best measured directly: either as the recoverable strain on sudden cessation of steady shear, or as the strain amplitude at which the storage and loss moduli cross over in an oscillatory amplitude sweep. Both these strain measures probe the same elastic-to-viscous transition and should be of similar order. Both measures are considerably smaller for particulate suspensions than their entangled polymer counterparts. Consequently, the same absolute strain-to-steady-state can mean very different things for different materials; a strain around magnitude ten, which still could be in the viscoelastic domain for an entangled polymer, could well be a genuine order-of-magnitude excess for a particulate suspension whose own elastic strain scale is significantly smaller. This is why Definition 1 is framed as a ratio of strain required to reach steady state *to the material's own viscoelastic strain*, rather than as a fixed threshold. It also suggests that measuring recoverable strain, or cross-over strain, more routinely than at present would allow future claims of thixotropy to be checked against an appropriate, material-specific baseline. With this understanding, Definition 1 offers a robust and practical operational criterion for thixotropy, requiring only a single well-specified experiment comprising a sufficiently large step-up in deformation rate or applied stress, ideally from a well-rested state, together with a suitable, material-appropriate reference strain.

Definition 1 applies equally when the controlling variable is stress rather than shear rate. Instead of imposing a step-up in shear rate, a material is subjected to constant stress, starting from a rested or rejuvenated state. Concurrently, accumulated strain is monitored before the material either reaches a new steady flowing state or, alternatively, shows no sign of doing so. This stress-controlled reading of Definition 1 recovers, as an important special case, the phenomenon widely known as viscosity bifurcation [41, 54, 55]. Below a critical stress, physical aging dominates: the viscosity and the associated relaxation time increase without bound, and no steady flowing state is ever reached. Above the critical stress, on the other hand mechanical rejuvenation dominates, and the material does reach a steady flowing state, but, consistent with Definition 1, only after a strain far exceeding what nonlinear viscoelasticity alone would require. The critical stress separating these two branches need not be introduced as an independent constitutive element. It instead emerges from the competition between structural buildup (physical aging) and stress-induced breakdown (mechanical rejuvenation), and it can itself increase with the rest time allowed prior to the imposition of stress [17, 55, 56]. This picture is consistent with the observation of Møller et al. [16] that much of the practical difficulty in measuring "the" yield stress of a material disappears once its thixotropic, history-dependent nature is properly accounted for. This threshold stress should not be confused with the yield stress of an ordinary yield-stress material which typically arrests flow after only a fraction of a strain unit. The requirement that a thixotropic material instead accumulate a strain far in excess of its viscoelastic reference strain, on the flowing branch, before the outcome becomes evident keeps Definition 1 consistent across its rate- and stress-controlled forms.

Under rate control, if the underlying constitutive equation is strictly non-monotonic, it can manifest macroscopically as shear banding, with the material separating into a flowing band and an arrested band whose relative widths are set by the lever rule [57]. We flagged this phenomenon earlier in this section as a primary reason to keep the imposed shear rates for a rate-controlled test of Definition 1 outside the non-monotonic region of the flow curve. We reiterate that a non-monotonic steady-state flow curve, or the shear banding it can produce, is not by itself a distinguishing signature of thixotropy, since both can also arise in nonlinear viscoelastic materials [58]. Therefore, they are mentioned here only as one possible macroscopic consequence of the aging–rejuvenation bifurcation that the stress-controlled form of Definition 1 captures directly.

We use the term "viscosity bifurcation" somewhat more broadly than in its original usage by Coussot and Bonn [41], which requires a non-monotonic constitutive curve and a discontinuous selection of shear rate under stress control. Here, even a near-plateau in the low-shear-rate region leading to a Herschel–Bulkley-type flow curve can produce an effective bifurcation, since incrementally imposed discrete values of stress (as it is not possible to experimentally impose continuous variation of stress with infinitesimal increment) still forces the system to select between an arrested and a flowing state. Indeed, Divoux et al. [59] showed experimentally that the stress-induced fluidization time of a simple yield-stress fluid diverges as the applied stress approaches the yield stress from above, providing a direct rationalization of the empirical Herschel–Bulkley law in exactly these terms. Non-monotonicity of the flow curve is, therefore, sufficient but not necessary for a viscosity bifurcation to occur.

It is important to comment on what distinguishes non-thixotropic yield stress materials from materials that show thixotropy as well as yield stress. Thixotropic materials undergo some viscous flow below the yield stress, while in conventional yield stress materials (that are approximated by a Bingham or Herschel–Bulkley type constitutive equation) the (small) elastic deformation present below the yield stress is completely recoverable. While the recoverability of strain below the yield stress could in principle be used to distinguish thixotropic yielding from ordinary viscoelastic-plastic yielding, in practice, not many researchers measure the recoverable strain, making this distinction difficult on that basis. Therefore, Definition 1 gives an important, easily observed, criterion for distinction, namely the amount of strain required for a material to reach steady state when subjected to stress above the yield stress. In the case of thixotropic materials, that strain is significantly larger than the recoverable strain, or the plausible maximum value of that recoverable strain (if, as often the case, it is not measured), while in the case of non-thixotropic conventional yield stress materials, it is of the same order as recoverable elastic strain or an estimate of it.

Furthermore, it is worth being explicit about one subtlety in incorporating viscosity bifurcation into Definition 1 in this fashion. The flowing branch of the bifurcation, on which a steady state is reached only after an anomalously large strain, is a direct realization of the criterion of Definition 1 itself. The aging branch, however, on which viscosity and relaxation time grow without bound and no steady state is ever reached, is not really a case of a "requirement of a large strain to achieve steady state", since no steady state exists to be reached on that branch. Interestingly this precise phenomenon is introduced below by the Definition 2, which captures directly the unbounded growth of the relaxation time with waiting time. Viscosity bifurcation is, therefore, best understood not as belonging exclusively to Definition 1, but as the phenomenon that joins Definitions 1 and 2 at a single, shared critical stress. Definition 1 monitors the strain as the steady state is approached on the flowing branch, while Definition 2, discussed below, governs the waiting-time-controlled divergence on the aging branch. We retain the discussion of viscosity bifurcation here, alongside Definition 1, because it is most naturally introduced as a stress-controlled counterpart to the rate-controlled criterion already given, but we return to its aging branch explicitly under Definition 2.

Both definitions above are framed in terms of responses that, however slowly, remain expressible through histories of shear rate, stress, or strain. A perspective distinct from these emerges when one asks instead whether the material shows an unbounded dependence on *the waiting time* between an initial and a final strain history, that does not saturate at any fixed relaxation timescale as the waiting time increases. This motivates a final, purely waiting-time-dependent definition, stated independently of the type of imposed initial and final deformation histories.

Definition 2: *Thixotropy is exhibited when, after cessation of an imposed shear or stress history, the material's mechanical relaxation time increases without bound during a subsequent rest period, as revealed by additional following rheological measurements*.

Definition 2 characterizes thixotropy through the time-dependent increase in the relaxation time ($\tau$) of a material, which can be at least partially reversed by resumption of sufficiently strong deformation field. The corresponding enhancement needs to be without bound with an increase in the waiting time $t_w$ elapsed since cessation of a prior shear or stress history. Note that $\tau$ sets the timescale over which the material's response retains memory of that history. We refer to this quantity throughout as the mechanical relaxation time, meaning specifically the characteristic timescale that governs the material's own rheological stress-relaxation or creep response, the same $\tau$ that appears in the relaxation modulus $G(t, -t_w t_w)$ or the creep compliance $J(t, -t_w t_w)$. Therefore, the unbounded growth of $\tau$ is revealed experimentally through these rheological response functions. These response functions depend explicitly on $t_w$, in addition to that on $t - t_w$, in violation of time-translational invariance and hence the Boltzmann superposition principle. The qualifier "without bound" prevents the transient response of a typical nonlinear viscoelastic material from being mistaken for Definition 2. In a standard nonlinear viscoelastic material, a flow imposed shortly after cessation of a preshear still carries some memory of that preshear, up to a timescale set by its own constant longest relaxation time. Once $t_w$ exceeds this fixed relaxation time, the response depends only on $t - t_w$, and no further growth of the relaxation time with $t_w$ occurs. A thixotropic material under Definition 2, by contrast, is one whose relaxation time itself keeps growing with $t_w$, so that the timescale over which stress relaxes continues to increase indefinitely, and the response function never settles into dependence on $t - t_w$ alone, notwithstanding how large $t_w$ is allowed to become. Although materials typically have many relaxation times, a precise definition of $\tau$ is not needed in our Definition, since a divergence in $\tau$ to at long $t_w$ will be seen in any average that encompasses this divergence. This will become clearer in the examples provided in Section III.

Non-rheological probes, such as dynamic light scattering, diffusing-wave spectroscopy, or microrheology based on tracer-particle motion, do not measure $\tau$ directly. They measure a structural or optical correlation time that reflects the same underlying microstructural relaxation. We admit such measurements as evidence for Definition 2 only when the characteristic decay time of a probe has been shown, or can reasonably be argued, to track the material's mechanical relaxation time for example through a generalized Stokes–Einstein relation connecting tracer mean-squared displacement to creep compliance [60], or through a demonstrated correlation with an independent measurement. The physical picture underlying Definition 2 is not confined to the colloidal and particulate systems examined explicitly in Section III. Its clearest antecedent is Struik's seminal work on physical aging in amorphous glassy polymers [7]. Struik was among the first who showed that for glassy polymers a rheological response function measured at a fixed elapsed time after a mechanical perturbation depends explicitly on the age of the material. Here, "age" is the time elapsed since it was quenched below its glass transition. According to Struik, this dependence of rheological response function does not saturate but continues indefinitely (or, more precisely, beyond any practical experimental window for deep quenches) as the material ages further. Now glassy polymeric systems are not conventionally regarded as "thixotropic," since the community studying them speaks the language of physical aging rather than thixotropy. Furthermore, the reference, zero-age, state in glassy polymers is established by quenching the material from above its glass transition temperature rather than by mechanically shear-melting it. Nonetheless, the phenomenon Struik identified, wherein the response function shows an explicit and unbounded dependence on waiting time, is exactly the signature Definition 2 requires. The additional requirement that the aging be reversible by flow is what separates thixotropic materials from glassy systems that can only be reset thermally. Struik's work laid the empirical and conceptual foundation for later, more general treatments.

Sollich et al. [61] generalized this picture into the theoretical framework of Soft Glassy Rheology (SGR), extending the notion of physical aging beyond molecular glasses to a broad class of soft materials, such as foams, emulsions, pastes, slurries, and colloidal glasses, etc. Sollich et al. represented these materials generically as an ensemble of mesoscopic elements trapped within a disordered energy landscape, without reference to any specific microstructure. Below an effective glass transition, the model does not reach the equilibrium state under quiescent conditions. Subsequent to sufficiently strong shearing flow (rather than by heating) which rejuvenates the material, during a waiting time with no flow the elements fall into ever deeper traps, increasing the material's effective relaxation time without bound, and this controls the response to subsequent deformation. This framework was later formalized into Struik-like waiting-time dependence within the SGR framework. Fielding *et al.* [62] showed that the resulting relaxation modulus and creep compliance depend on the waiting time in addition to the elapsed time since the probe was applied, violating time-translational invariance in precisely the manner Definition 2 requires. Because this framework, like the definitions proposed here, is deliberately agnostic to the specific microstructural origin of the aging it describes, it offers independent theoretical support for Definition 2, and brings the broad class of soft glassy materials, which age under quiescent conditions and mechanically rejuvenate under an applied deformation field, within the scope of thixotropy as defined in Definition 2. Of course, we are not the first to bring soft glassy materials under the purview of thixotropy and Mewis and Wagner explicitly make this connection in their seminal work [4].

Again, we are not insisting that rheologists studying either hard or soft glassy materials adopt the term “thixotropy” in place of their well-established terminology, but we observe, for clarification, that the “aging” rheological phenomena observed in those fields is frequently called “thixotropy” when present in hard-particle suspensions and other materials.

## Discussion

The three definitions divide naturally into two groups by how they anchor thixotropic behavior. Definitions 0 and 1 tie thixotropy to a history of imposed deformation: a rheological property, viscosity, relaxation time, or yield stress, decreases under sustained shear or stress applied at a sufficiently high rate or level. Definition 0 further associates thixotropy with recovery of the same property during rest or under a weaker field. Definition 1, on the other hand, in its stress-controlled, bifurcating form, applies to the flowing branch above the critical stress. Within this group, the definitions differ in which property is emphasized, whether shear rate or stress is the controlling variable, and whether accumulated strain is invoked explicitly to exclude viscoelastic contributions. Definition 2 differs from this group in that, in principle, it makes no reference to a specific deformation history but instead identifies thixotropy by the unbounded growth of the material's relaxation time with waiting time; but with a condition that such growth must get reversed under application of sufficiently strong deformation field. For Definition 2, the applied field at different waiting times could be stress- or strain-controlled; it is the unbounded growth of the relaxation time with waiting time that matters, not the particular protocol used to assess the growth of relaxation time. Such growth is revealed by a response function that explicitly depends on waiting time. An ordinary viscoelastic liquid, whose memory of a prior perturbation fades once the waiting time exceeds its maximum relaxation time, does not show such an unbounded growth, while an ordinary viscoelastic solid relaxes to its final stress after deformation ceases in a time governed by a finite longest relaxation time. We recognize that polymer liquid and solids commonly recognized as “viscoelastic,” can have very long relaxation times that can change with time, and so there is some judgement required in determining how long one must wait before inferring that the relaxation time grows “without bound,” when one applies the above criteria for “thixotropy.”

Considered individually, the three definitions represent distinct but compatible perspectives on thixotropy, and each presents its own practical difficulty. Definition 0 offers the cleanest conceptual baseline but is ambiguous about how much elastic contribution can be tolerated before a response ceases to be "thixotropic." This is a real difficulty, since most materials conventionally called thixotropic do show appreciable linear elasticity in small-amplitude oscillatory shear, often with a storage modulus exceeding the loss modulus. Definition 1 addresses this in practice by locating the boundary with viscoelasticity in the magnitude of accumulated strain relative to the elastic strain rather than in the magnitude of elasticity itself, and by phrasing thixotropy directly in terms of what is actually measured: the gradual fall of viscosity, following a sufficiently large increase in deformation rate or applied stress, over strain far exceeding what viscoelasticity alone would require. In its stress-controlled form, Definition 1 also supplies the conceptual rationale behind viscosity bifurcation, recasting the large-strain criterion as a competition between physical aging below a critical stress and mechanical rejuvenation above it. Definition 2, on the other hand, isolates thixotropy through the unbounded growth of the material's own relaxation time with waiting time. It offers a criterion independent of whichever deformation protocol, stress-controlled or rate-controlled, is used to reveal that growth, and it extends naturally to the aging dynamics of soft glassy materials described by the SGR framework. Taken together, the three definitions capture the different guises in which thixotropy has historically been reported, while excluding the nonlinear viscoelastic response.

This last exclusion is only as sharp as the boundary drawn between a "viscoelastic" and a "thixotropic" response. Definitions 0 and 1 both invoke viscoelasticity as the behavior to be departed from, but unless that boundary is made concrete, it risks being stretched to cover any response involving stress relaxation, however slow, leaving nothing available to be classified as thixotropy. The boundary thus needs references against which departures from viscoelasticity can be measured. To help identify such boundaries, we take "quintessential" viscoelasticity to be the behavior of simple, non-associating, high-molecular-weight homopolymer solutions or melts, and "quintessential" thixotropy to be the behavior of simple suspensions of spheroidal particles undergoing gelation through van der Waals or depletion forces. Many real materials such as associating polymers, non-spherical gelling particles, soft and hard glasses, emulsions, blood, liquid crystals, crystallizing systems, and mixtures of these, sit somewhere between these two “quintessential” cases and can display characteristics of both; the two reference materials given above are what allow individual rheological features of such intermediate systems to be assigned to one limit or the other.

Anchored at each limit to the behavior of the two quintessential materials, two operational metrics follow — one for the strain-based definitions, and one for the waiting-time-based one. For Definitions 0 and 1, the relevant metric is recoverable strain. Simple homopolymer solutions and melts reach steady state over a strain within an order of magnitude of their maximum recoverable strain, which, as per general experience, rarely exceeds unity in a colloidal suspension. Since recoverable strain can in principle be measured for any material, it serves as the strain scale above which the "negligible" elasticity required by Definition 0, and the strain-driven approach to steady state, under either rate or stress control, together with the associated critical-stress bifurcation, that together define Definition 1, can all be recognized as distinctively thixotropic rather than as the result of long nonlinear viscoelastic transients. For Definition 2, the analogous metric is primarily the time: the unbounded growth of the material's own relaxation time with waiting time can be attributed to thixotropy. On the other hand, a standard non-linear viscoelastic material possesses a constant longest relaxation time or at least a bounded one, and beyond that, the response of a material loses any waiting time dependence. Without a cutoff of this kind, very slow colloidal rearrangements that relax free energy and therefore always leave some trace, however weak, in the stress would qualify as "viscoelastic" no matter how long they took or how much strain they were sensitive to. Excluding this possibility is what keeps the three definitions operationally robust.

We note that some may feel that, since a “thixotropic” response is, in terms of fundamental physics, actually just an extreme of viscoelasticity, there is no need to introduce “thixotropy” as a separate concept. However, as is widely recognized, rheological categories are oriented towards pragmatic applications, rather than fundamental physics, and widely used concepts such as “Newtonian fluids,” “shear thinning inelastic fluids," and even "yield stress" fluids are idealizations that are highly useful in practice [36], however non-rigorous they may be from the perspective of pure physics. The rigor needed for a rheological category is in its mathematical expression, not in the perfection of its realization in actual materials.

We do not claim that these definitions are the only ones that characterize thixotropic behavior, but we do suggest that not all rheological properties that seem to be consistent with thixotropy are suitable as *defining* characteristics, rather than *diagnostic* ones. Many properties, including size and shape of hysteresis loops in up and down shear ramps, shapes of nonlinear Lissajous curves in nonlinear oscillatory shear, and other time- and shear-sensitive nonlinear responses can correlate with thixotropy. Having a limited number of clear definitions, however, should help clarify communication, and we recommend that any additional definitions be given careful thought and be tested for robustness against experimental literature, as we do in the next section. Properties for which distinctions between thixotropy and viscoelasticity are less clear might better be left as diagnostic characteristics, rather than defining ones, since there can be exceptions to diagnostic behavior that cannot be tolerated in defining characteristics without introducing confusion.

While the mathematical form of existing thixotropic constitutive equations has guided our discussion, the above definitions, except possibly Definition 0, remain deliberately phenomenological throughout. None prescribes a specific constitutive model or microstructural picture, so "thixotropy," depending on which definition is invoked, may in principle apply to glasses, liquid crystals, metals, pastes, or any other material with a complex, delayed response to shear history, even where an established, field-specific terminology might otherwise discourage the label. Given this latitude, we recommend that rheologists state explicitly which definition they intend, whether one of the above, or some other definition, when describing a material or its response as "thixotropic," so as to reduce ambiguity in the literature. The next section illustrates the application of each of the three definitions given above to real materials.

**III. Experimental realizations of the proposed definitions:**

In this section, we discuss various experimental observations on thixotropy from the literature. We note that definitions expressed in continuum-mechanical terms (e.g., viscosity, relaxation time, yield stress, accumulated strain) are useful only when they can be verified against experimental data. In addition, phrases like "much larger strain" or "changes without bound" need to be transformed into measurable thresholds that readers can use without additional speculation. This section has two purposes. The first is to collect and categorize representative examples from the literature for which Definitions 1 and 2 can be applied, each drawn from a distinct experimental protocol. The second is to attempt to associate an explicit, quantitative criterion with each definition, so that the term "thixotropy" becomes a verifiable classification rather than simply the label the original authors of the respective papers happened to use. In the following discussion, definition 0 does not appear because, as mentioned in Section II, it describes an idealized, inelastic limit that is best understood as a modeling reference state rather than a response expected of any real physical material. Owing to its idealized nature, it has been labelled as Definition 0.

The selection of the examples discussed here was guided by a set of qualitative criteria, as explicitly stated before the data were discussed. First, we only include materials whose time dependence falls within the scope of what is considered a rheological property. Thus, the time dependence must be attributed, with reasonable confidence, to reversible, shear-induced *microstructural* rearrangement, rather than to *chemical* or *macroscopic* changes in the material. Consequently, among the cases that we exclude (but not limited to these) are: (i) Systems undergoing irreversible chemical reactions on the timescale of the measurement. An example is cement paste during early hydration, where the time dependence of rheology reflects chemical conversion rather than reversible structural kinetics [63]. (ii) Systems that develop macroscopic inhomogeneities such as shear bands. These include lamellar gels [64]; see also the broader review by Divoux et al. [58], wherein the apparent rheological time dependence originates from band formation rather than an intrinsic thixotropic phenomenon. (We remind the reader that the above should not be taken to imply that thixotropic systems do not exhibit shear banding.) (iii) Liquid-crystalline materials where distortions of the director field on a macroscopic scale (i.e., order tens of microns or larger) produce time-dependent rheological effects. And (iv) non-stabilized multiphase liquid–liquid or liquid–air systems such as some emulsions or foams, whose dynamics are governed by unbounded droplet or bubble coarsening. These cases are distinguished from other emulsion systems (such as mayonnaise [65, 66]) that are potentially thixotropic if their liquid or gas domains respond to flow, but remain microscopic for indefinitely long periods of time without separating into macroscopically separate phases. We again specify that rheological responses to changes that are either chemically irreversible, such as covalent bond formation, or involve macroscopic length scales, such as unbounded emulsion or foam coarsening, are not considered under the scope of thixotropy, or of any *rheological* material property.

Other exclusions from consideration here, although not necessarily excluded from future consideration, are cases in which details are lacking from the publication, such as a clearly identified strain, waiting time, or sufficiently prolonged shearing, the absence of which prevents confident classification under our definitions. Thus, we have deliberately confined the selection to cases we regard as "clear-cut," leaving aside for future consideration materials and data sets that are plausibly, but not unambiguously, thixotropic. Finally, the experimental examples we examine are not exhaustive but representative, covering a versatile range of materials and microstructures. In the following discussion, we consider the proposed definitions and relate them to experimental reports in the literature.

### III.A. Definition 1

Definition 1 proposes that, following a sufficiently large increase in shear rate or applied stress, the strain required for a chosen rheological property, such as viscosity, stress, or storage modulus, to attain its new steady value must exceed significantly (say at least a factor of ten) its expected recoverable elastic strain. For ordinary particulate suspensions, this viscoelastic reference strain is expected to be no larger than unity, as mentioned in Section II. Thus, for a particulate system, whenever the actual strain required to attain a steady state exceeds tens of strain units, the response is already comfortably within the thixotropic domain according to this criterion. In the examples given below, we begin with the rate-controlled form of this criterion (A.1). The corresponding stress-controlled form, and the viscosity bifurcation it recovers as a special case, are taken up later in this section (A.2).

### III.A.1 Rate-controlled flow field

The six examples given below demonstrate this operational criterion over a wide range of materials and experimental techniques. The first example we consider is a cellulose nanocrystal (CNC) gel studied by Jia et al. [67]. The authors investigated a gel prepared by dispersing 5 wt% cellulose nanocrystals, which are rod-shaped particles roughly 3–25 nm in diameter and a few hundred nanometers long, in water containing 20 mM NaCl. The added electrolyte screens the electrostatic repulsion between the negatively charged rods, allowing the attractive component of the interparticle force to dominate, leading to a weak, space-spanning aggregated network. Jia et al. [67] employed twelve fixed shear rates spanning 0.1–2000/s. A freshly loaded sample was sheared for 10 min and rested for an additional 10 min. Subsequently it was sheared continuously for a further 10 min (600 s) at a specified shear rate.

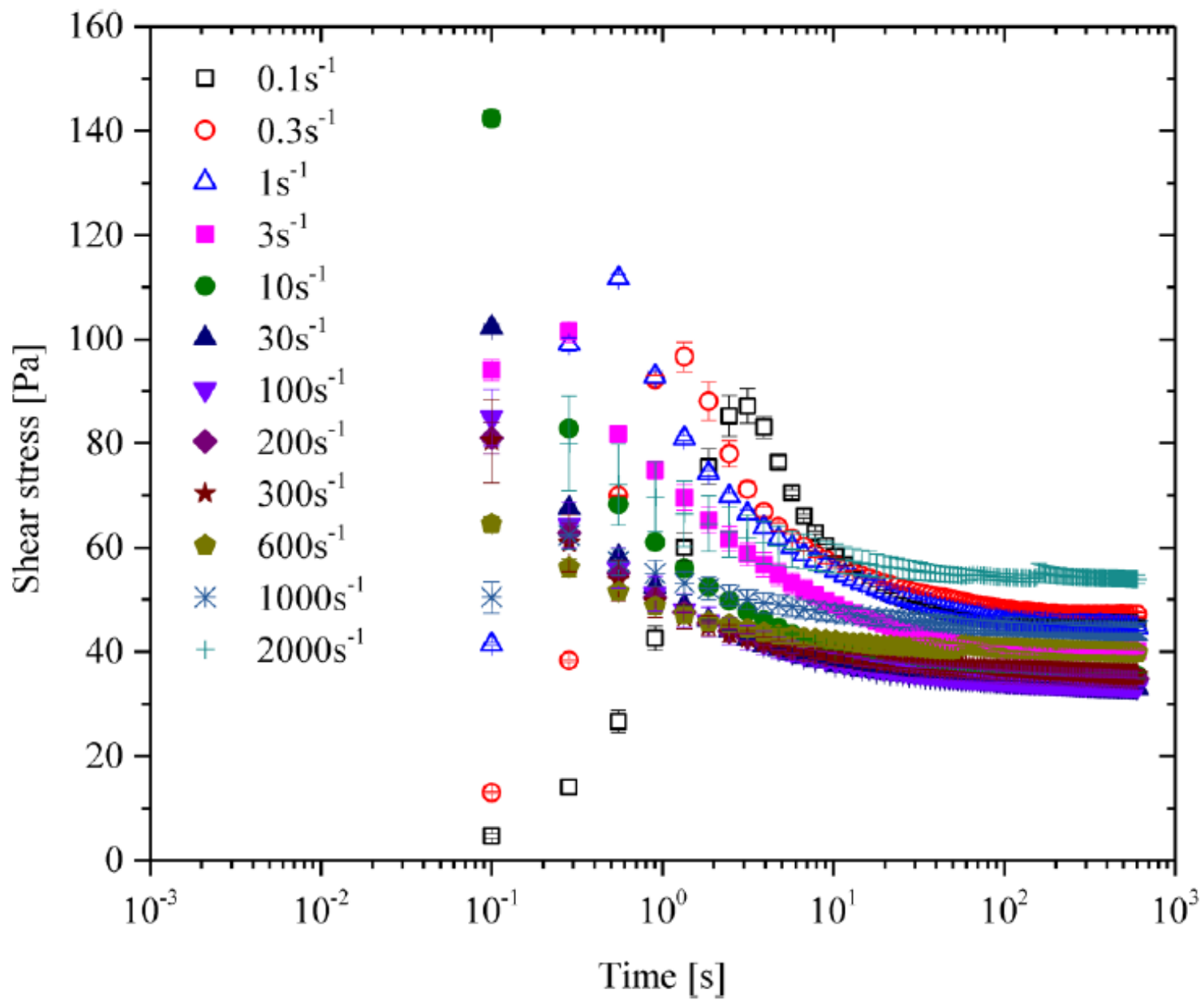

**Figure 1.** Transient shear stress as a function of time for a 5 wt % cellulose nanocrystal gel (containing 20 mM NaCl) following a step increase in shear rate from a rested state to the constant values given in the legend (0.1–2000/s). Each curve is an average over several repeats on a freshly loaded sample that is rested for 10 min and then sheared for 10 min; a repeatable, steady stress value is reached at each rate shown. The steady-state

flow curve of this system is non-monotonic in the domain 0.3–30/s. Adapted, with permission, from Jia et al. [67].

Jia et al. claim that the material possesses non-monotonic flow curve with shear rates in the domain 0.3–30/s belonging to the region associated with negative slope of shear stress – shear rate curve. Furthermore, authors report noticeable wall slip at the low shear rates. Therefore, in Fig. 1, we do not consider the data belonging to shear rates ≤30/s. It can be seen that at the shear rates 100/s and above, which lie on the monotonic branch of the flow curve, the transient stress curves do reach a steady-state plateau, but the corresponding accumulated strain runs into the hundreds and thousands of strain units. Since this is much larger than can be attributed to a viscoelastic microstructure, especially for a suspension whose elastic or recoverable strain is likely no higher than order unity, this example obviously exceeds the threshold specified in Definition 1.

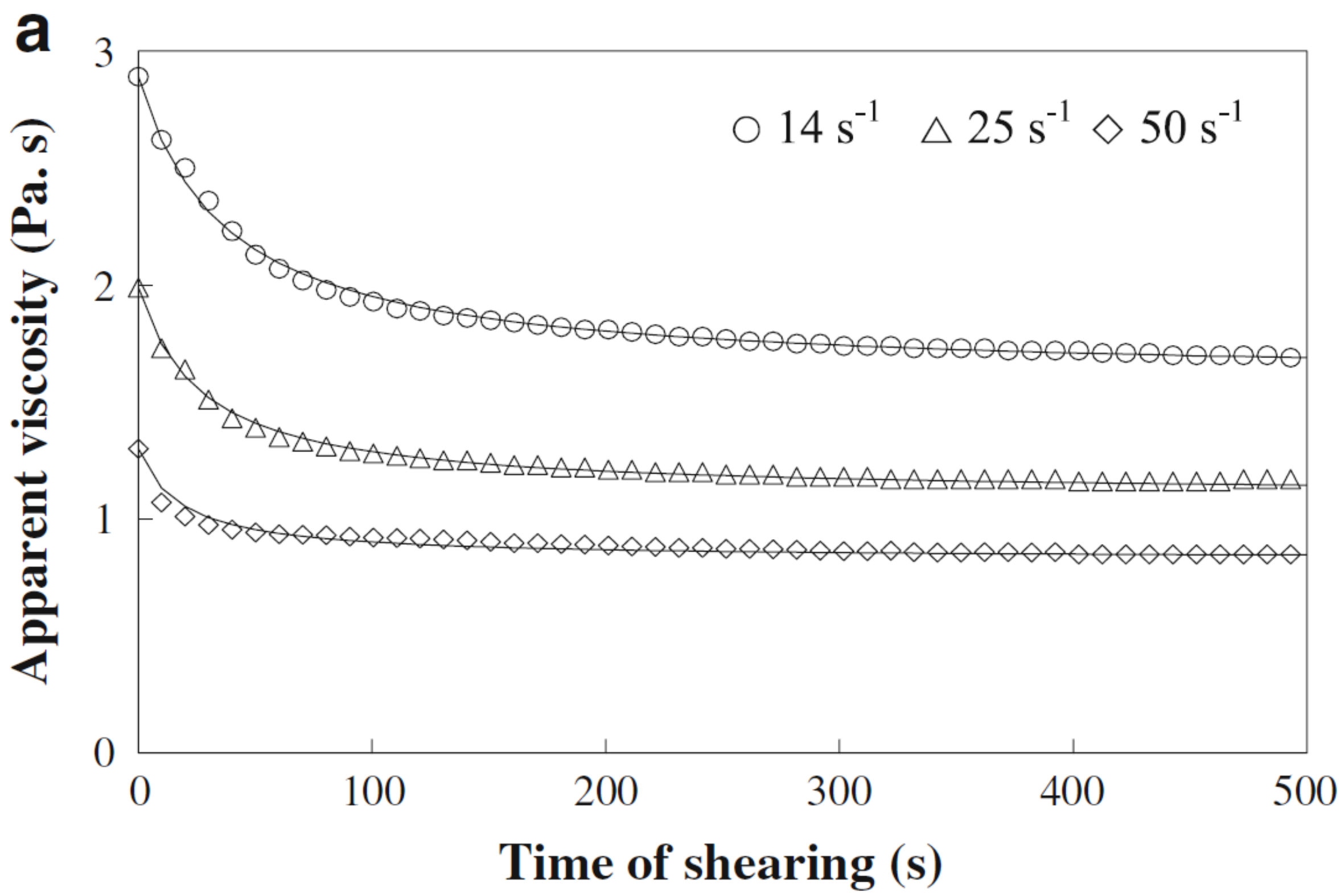

**Figure 2.** Viscosity as a function of shearing time for a 3% (w/w) Alyssum homolocarpum seed gum solution at 25°C. The corresponding shear rates are mentioned in the legend. The solid lines represent the fits of a second-order structural kinetic model employed by the authors. Adapted, with permission, from Koocheki and Razavi [68].

Next, we consider an aqueous solution of the mucilaginous gum extracted from *Alyssum homolocarpum* seeds, a hydrocolloid used as a thickening agent in food products, investigated by Koocheki and Razavi [68]. In their work, a 3% (w/w) solution, equilibrated at 25°C, was subjected to constant shear rates of 14, 25, and 50/s, and the corresponding evolution of viscosity was measured over 500 s. It can be seen in Fig. 2 that, at each applied shear rate, the apparent viscosity decreases sharply over the first 50 seconds or so and then continues to decay gradually, eventually approaching a near-constant value beyond 100 -300 s. Taking the lowest rate, 14 $s^{-1}$, as an example, supposing that the viscosity levels after 100 s, the strain associated with reaching an operationally steady value is around 1400 strain units. Even the highest rate of 50/s, which approaches the steady state comparatively sooner (≈ 60 s), still corresponds to a few thousand strain units. Since no viscoelastic reference strain can compare with such a magnitude, the seed gum solution can be unambiguously placed within the scope of Definition 1 to be thixotropic. This example shows that thixotropic behavior is not confined to synthetic colloidal particles but also extends to naturally occurring polysaccharide networks widely used in the food industry.

Ogunwale et al. [8] studied a model waxy crude oil, "SIGMA," prepared by dissolving 10 wt % paraffin wax in a heavy mineral oil, the same formulation used in earlier studies of wax gelation [9]. On cooling below its wax-appearance temperature under continued shearing, dissolved wax molecules crystallize into platelets that interlock into a space-filling network, converting the oil from a simple liquid into a soft solid-like paste with a measurable yield stress. In an experiment, the oil was cooled from 60°C to a final temperature of 26°C at a controlled cooling rate while being continuously sheared at 1/s. Ogunwale et al. [8] continued shearing at this same rate for a further 48 h at 26°C simultaneously recording the shear stress.

Fig. 3 shows that under continuous shearing at 1/s the stress decreases by more than an order of magnitude, from around 13 Pa to below 1 Pa, over the full 48 h duration. Interestingly, even after two days of continuous shearing, the decay does not visibly level off into a steady-state plateau. Since the shear rate is held fixed at 1/s throughout, the accumulated strain by the end of the test is over $10^5$ strain units, which is several orders of magnitude beyond what any viscoelastic response would require. Even a conservative reading of the data, taking only the more rapid initial fall over the first several hours as the relevant approach to an apparent, if not final, steady value, still implies a strain of tens of thousands of units. From the point of view of Definition 1, this example unambiguously shows that the studied model waxy crude oil is thixotropic by this criterion.

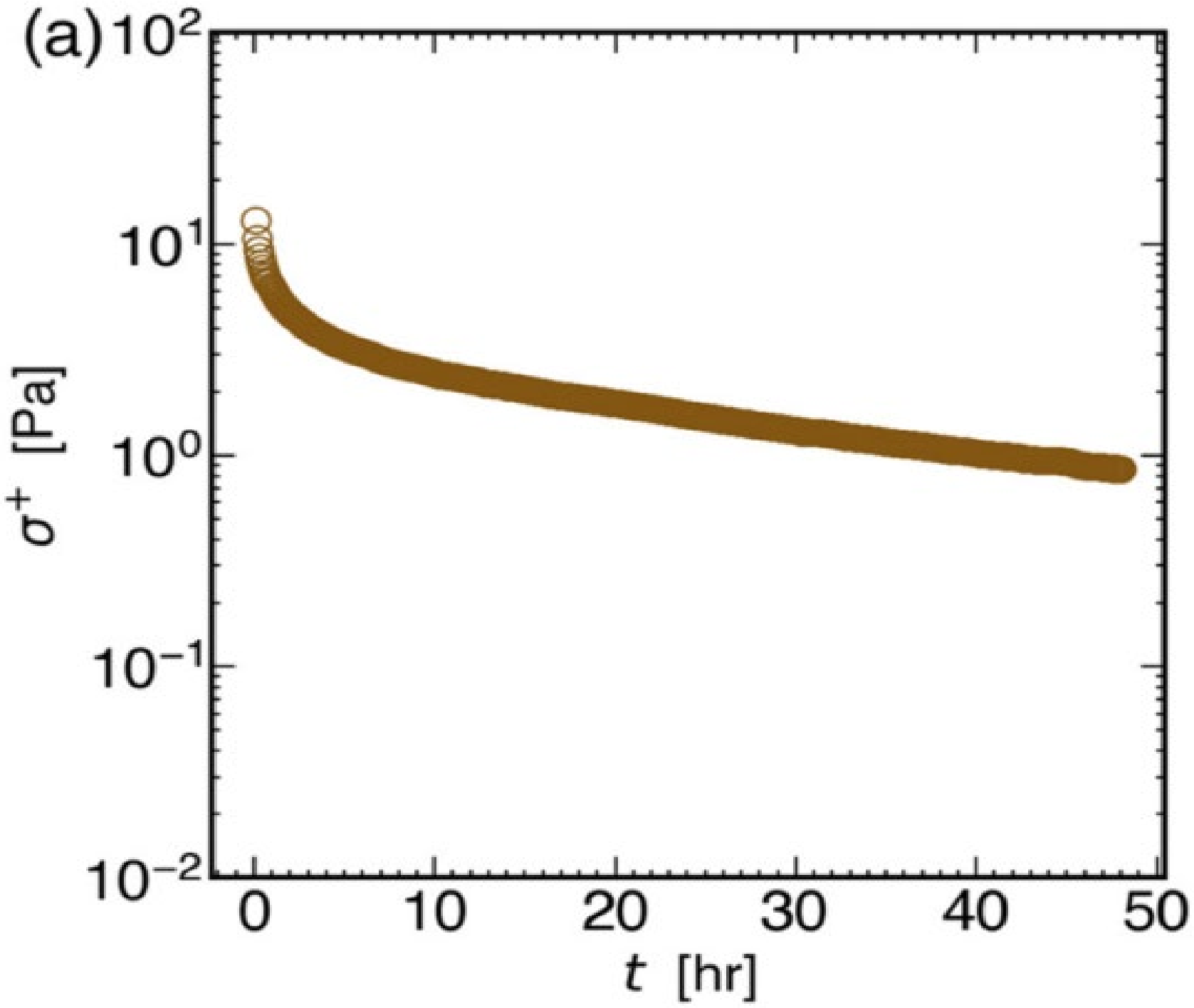

**Figure 3.** Shear stress as a function of time after attainment of the final temperature for the model waxy crude oil "SIGMA" (10 wt % paraffin wax in mineral oil), cooled from temperatures above the wax appearance temperature, under continuous shearing at 1/s. Adapted, with permission, from Ogunwale et al. [8].

Paszkowski and Olsztyńska-Janus [69] studied a lithium-soap-thickened lubricating grease, a three-dimensional network of lithium 12-hydroxystearate fibers immersed in mineral base oil. They used a two-stage protocol: in the first stage, a grease sample was sheared continuously at a fixed rate for 1 h, then left to relax undisturbed for 24 h; in the second stage, it was re-sheared at the same rate for a second 1 h period. This protocol allowed them to assess the extent of microstructural recovery during the rest period. The corresponding time evolution of stress for both protocols is shown in Fig. 4. Under continuous application of a shear rate of 8.1/s, the stress decreases from approximately 1600 Pa to 600–700 Pa during stage 1 (60 min). It can be seen that a gentle downward slope in stress persists even at the end of the hour, without reaching a plateau.

Sixty minutes of continuous shearing at 8.1 $s^{-1}$ corresponds to an accumulated strain of over $10^4$ strain units, which is several orders of magnitude beyond the strain of order unity expected of a viscoelastic material. However, the recovery observed for this material over 24 h of rest is rather very small. In the second stage the stress again decays without reaching a plateau over a comparable strain. This grease therefore satisfies the strain criterion of Definition 1.

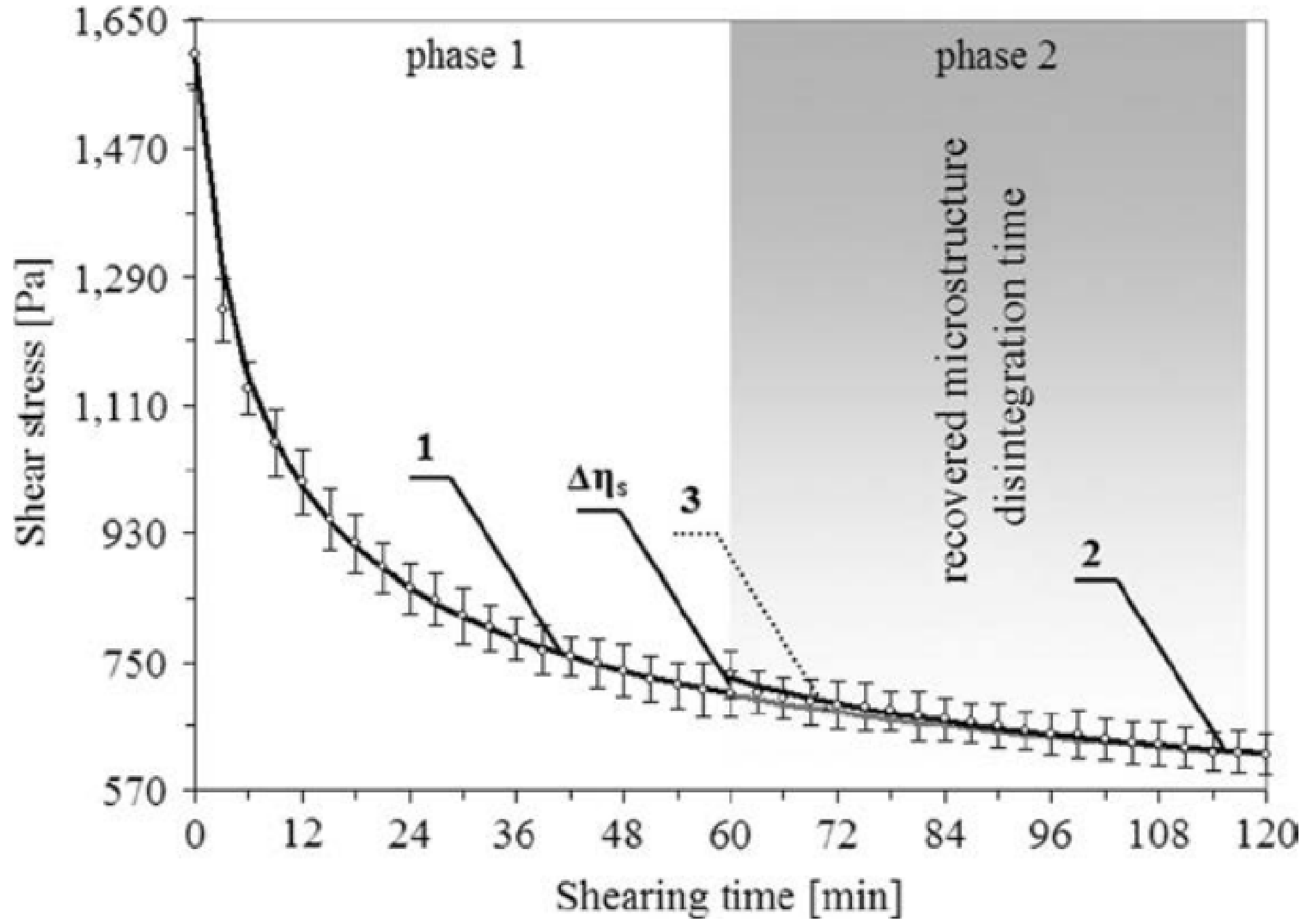

**Figure 4.** Shear stress as a function of shearing time for a lithium-thickened lubricating grease at a constant shear rate of 8.1/s. Phase 1 represents the first 60 min of shearing on the freshly kneaded sample, while phase 2 describes a following additional 60 min period of shearing after 24 h of undisturbed rest. The behavior is shown here on a continuous time axis for ease of comparison, though the two stages are separated in the actual protocol by the 24 h of rest period. Please refer to the original paper for details about markings 1, 2 and 3. Adapted, with permission, from Paszkowski and Olsztyńska-Janus [69].

Fumed silica and carbon black dispersions are considered to be canonical thixotropic materials in the literature [70]. Dullaert and Mewis [71] studied these dispersions wherein the corresponding attractive particles were suspended in Newtonian fluids: 2.9 vol % fumed silica particles in a paraffin oil/polyisobutylene mixture, and 3.23 vol % carbon black particles in a naphthenic oil. They subjected each dispersion to a step increase in shear rate, from an initial rate of 0.1/s up to final rates of 0.25 to 5/s, and recorded the resulting stress transient.

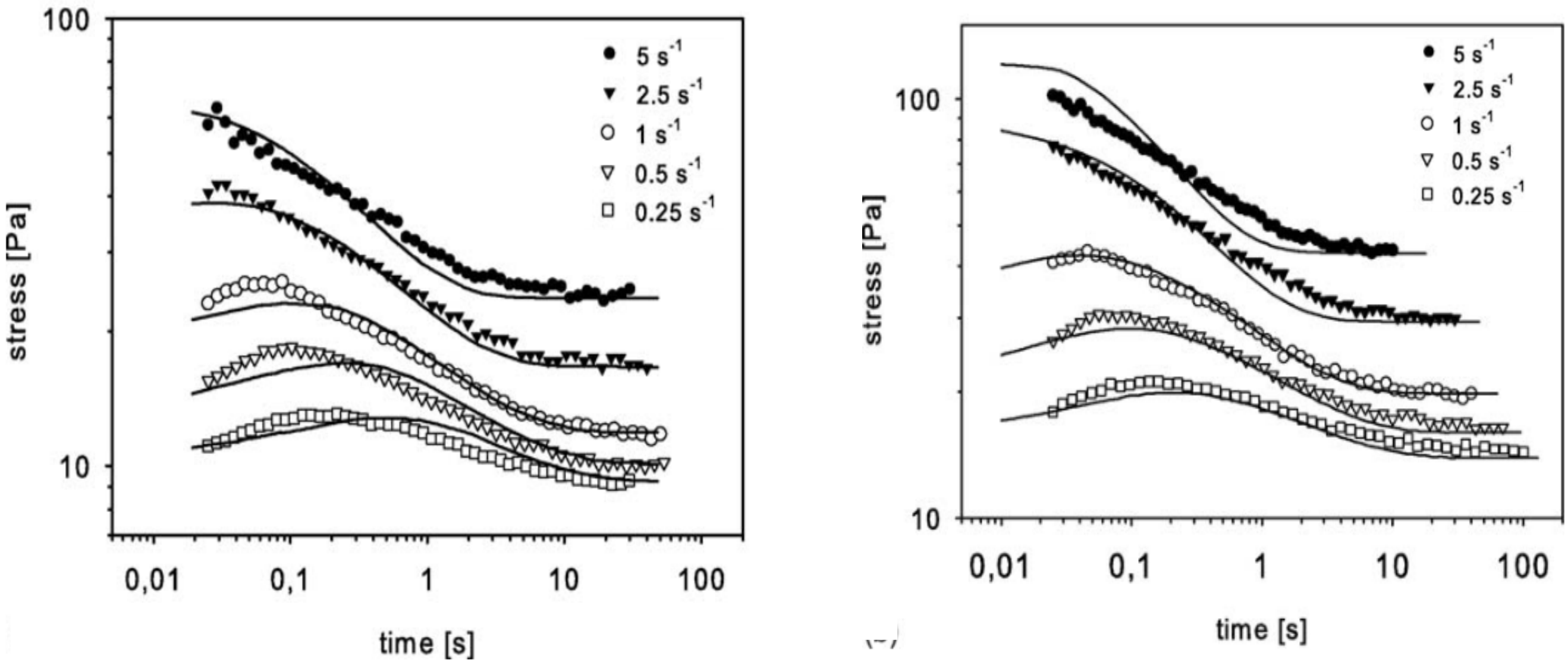

**Figure 5.** Transient shear stress as a function of time following a step increase in shear rate from an initial value of 0.1 $s^{-1}$ to the final values marked in the legend (0.25–5 $s^{-1}$), for (a) a 2.9 vol % fumed silica dispersion and (b) a 3.23 vol % carbon black dispersion, both in Newtonian suspending oils. Solid lines are predictions of the structural kinetics model of Dullaert and Mewis [71]; symbols are experimental data. Adapted, with permission, from Dullaert and Mewis [71].

As shown in Fig. 5, both dispersions show a brief stress overshoot over a short period of less than a second after the step, reflecting the elastic stretching of the particle flocs. Subsequently, the stress slowly decreases over one to two decades of time towards a steady value. Dullaert and Mewis [71] obtained the critical elastic strain ($\gamma_c$) that the flocs can sustain before yielding as the ratio of the extrapolated apparent yield stress ($\sigma_{y,0}$) to the low-frequency plateau storage modulus ($G_0$), $\gamma_c = \sigma_{y,0}/G_0$. For the silica dispersion, they obtained $\gamma_c$ ≈0.015 while for the carbon black dispersion, they estimated $\gamma_c$ ≈0.010, which one can take to be the material's viscoelastic strain scale. While the strain at which the stress transients shown in Fig. 5 attains a plateau is around ten strain units, which would not be considered large for a viscoelastic polymer, this exceeds $\gamma_c$ by three orders of magnitude and thus is far beyond the tenfold margin that Definition 1 requires. The smallest step (0.1 to 0.25/s) is less than one decade, so the criterion is best applied to the larger steps (final rates of 1/s and above). This picture was later corroborated directly by Choi et al. [72], who applied stress-controlled recovery rheology to fumed silica system during a step-down in shear rate. They showed that the transient elastic stress closely tracks the total stress at short times before the two diverge. This observation confirms that elasticity, though present, cannot by itself account for the much larger strain required to reach the new steady flow state. Carbon black dispersions have also been used extensively to develop and validate structural-kinetics constitutive models under large-amplitude oscillatory and reversing shear flows [73].

Another canonical example of a thixotropic system is an aqueous dispersion of Laponite® clay. Agarwal et al. [13] studied a 407 days old 3.5 wt% aqueous dispersion of Laponite® RD, a synthetic clay composed of charged, disc-shaped particles approximately 25–30 nm in diameter and 1 nm in thickness. This system is recognized for forming a space-spanning percolated network, resembling a "house-of-cards" structure [74]. In the study, the sample was first subjected to a preshear at a rate of 50/s for 300 seconds to eliminate any prior shear history. After the preshear, the sample was allowed to rest under quiescent conditions with zero shear for various waiting times (600, 1800, or 3600 seconds). Following the resting period, a step increase in shear rate to 3/s was applied, and the corresponding stress was recorded.

As illustrated in Fig. 6, the transient stress exhibits an initial overshoot lasting less than a second. The height of this overshoot increases systematically with the length of the preceding rest period. After this peak, the stress gradually decays over approximately three decades of time until the curves for different waiting times converge near the end of the 1800 s observation window following the step change. At the applied rate of 3/s, this results in an accumulated strain of several thousand strain units before reaching a steady state, which is significantly greater than what would be expected from a typical viscoelastic reference strain value. The systematic increase in overshoot with waiting time is particularly relevant to the waiting-time-based approach discussed in Definition 2 later in this section. Notably, the criterion outlined in Definition 1 is met regardless of the chosen waiting time. After the step increase in rate, the strain needed to achieve a new steady stress is much larger than any deformation that can be attributed to viscoelasticity, confirming that the laponite dispersion, along with the other five systems referenced, clearly aligns with the criteria set forth in Definition 1.

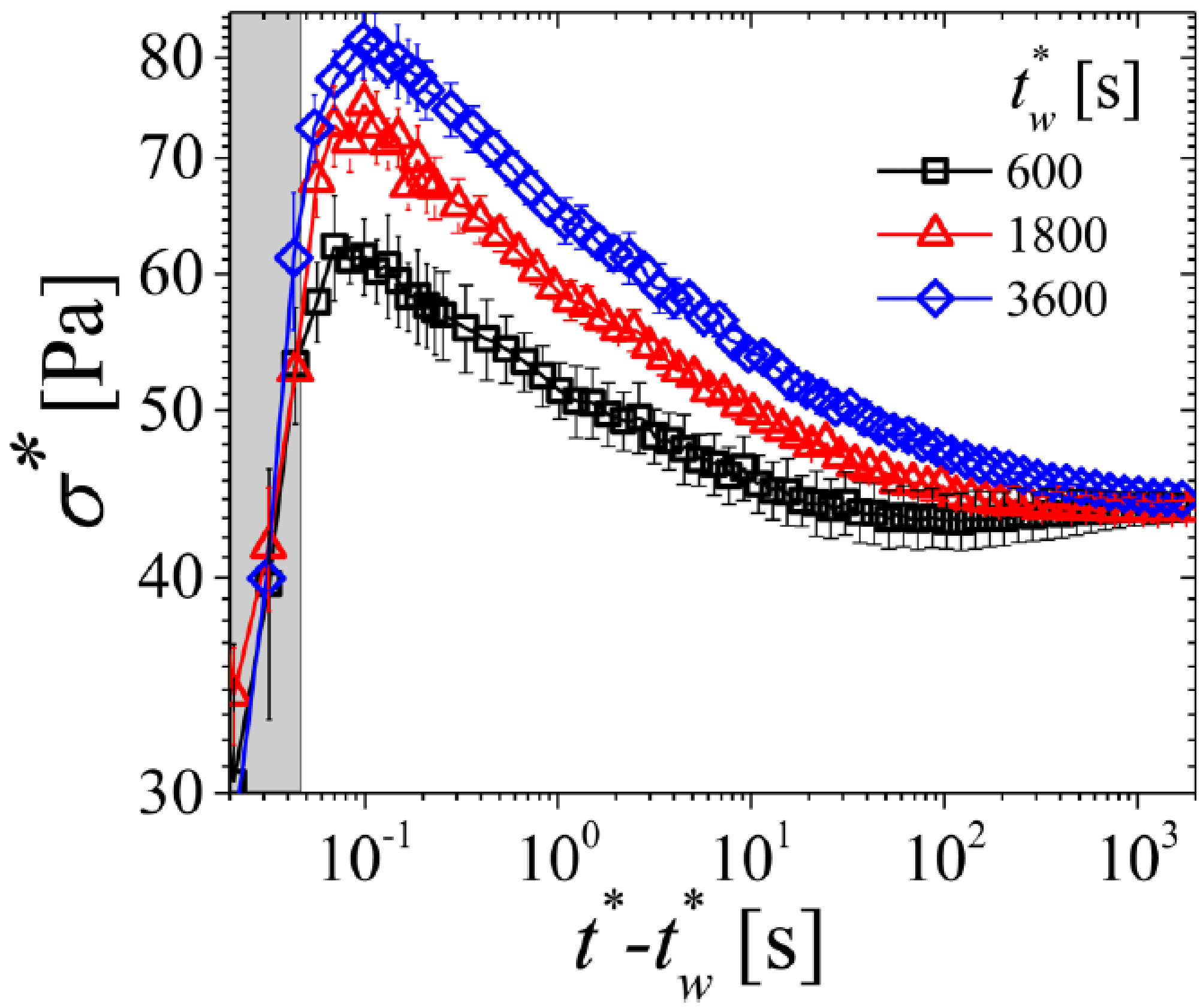

**Figure 6.** Transient shear stress versus time following start-up of shear at shear rate 3/s, applied after a period of undisturbed rest of 600, 1800, or 3600 s (legend), for a 3.5 wt % aqueous laponite clay dispersion . Data below 0.05 s (shaded) are affected by instrument inertia. Adapted, with permission, from Agarwal et al. [13].

The systems examined above to support the rate-controlled form of Definition 1 include a variety of microstructures, such as food hydrocolloids, a petroleum wax network, a soap-thickened lubricant, and three classes of colloidal particulate dispersions. The studies discussed here come from completely different practical applications, including rheology, food science, petroleum engineering, tribology, and colloid physics. All six examples significantly exceed the criteria outlined in Definition 1, not only meeting the required factor of 10 but typically exceeding it by two to five orders of magnitude. This consistency across a diverse range of materials is precisely the type of evidence that Definition 1 aims to organize.

### III. A.2 Stress controlled flow field

We now turn to examples of the stress-controlled form of Definition 1. Here, the imposed deformation is a constant applied stress rather than a shear rate. As in Section III.A, this reflects the underlying competition between structural buildup (also referred to as physical aging) and breakdown (also termed mechanical rejuvenation), except now with stress rather than strain rate as the control parameter. Under stress control, since now the strain is induced quantity, this competition takes on a sharper signature than under rate control. Rather than a smoothly decaying transient approaching a single steady value, as seen across the examples discussed above, the long-term behavior of a material under a constant applied stress depends critically on whether that stress lies above or below a threshold value. Below this threshold, the viscosity and the associated relaxation time increase without any apparent bound, driving the material toward an arrested, solid-like state where flow ceases. Above the threshold stress, on the other hand, the viscosity decreases and then levels off to a finite, steady value, allowing the material to flow continuously. This qualitative split in long-term behavior, widely termed viscosity bifurcation, is one of the most striking rheological signatures of thixotropic materials, which is captured by the stress-controlled form of Definition 1. We note, however, that simple viscous or viscoelastic yield-stress materials also show a bifurcation, with continuous flow occurring above the yield stress and small or negligible elastic deformation occurring below it. Hence, as in the rate-controlled case above, the issue of strain emerges as the distinguishing characteristic between thixotropic and "ordinary" yield-stress behavior.

The term viscosity bifurcation, and much of the experimental methodology used to characterize it, originates in the seminal work of Coussot et al. [41]. They subjected a bentonite clay suspension to a step stress after a period of rest and observed that the material either *eventually* ceases to flow below a critical stress or attains a finite steady shear rate above it. The corresponding critical stress has been reported to increase with the duration of the preceding rest [41]. As already noted in Section II, our usage of the term here is deliberately broader than the original Coussot–Bonn criterion. According to the Coussot–Bonn proposal, a viscosity bifurcation necessarily requires the underlying steady-state flow curve to be non-monotonic, so that the imposed stress in a discontinuous fashion selects the stable value of shear rate above minima. In this study, we focus solely on the operational, and more experimentally accessible, signature of the phenomenon: a threshold stress that separates an aging, arresting branch from a rejuvenating, continuously flowing branch. Since experimentally, stress can take only discrete values, this distinction holds true whether the underlying steady-flow curve is strictly non-monotonic or simply exhibits a low-shear-rate plateau, as in the Bingham or Herschel–Bulkley model. Consistent with the historical progression of this concept, our initial example is the same as Coussot et al. [41] 's original system: a bentonite clay suspension.

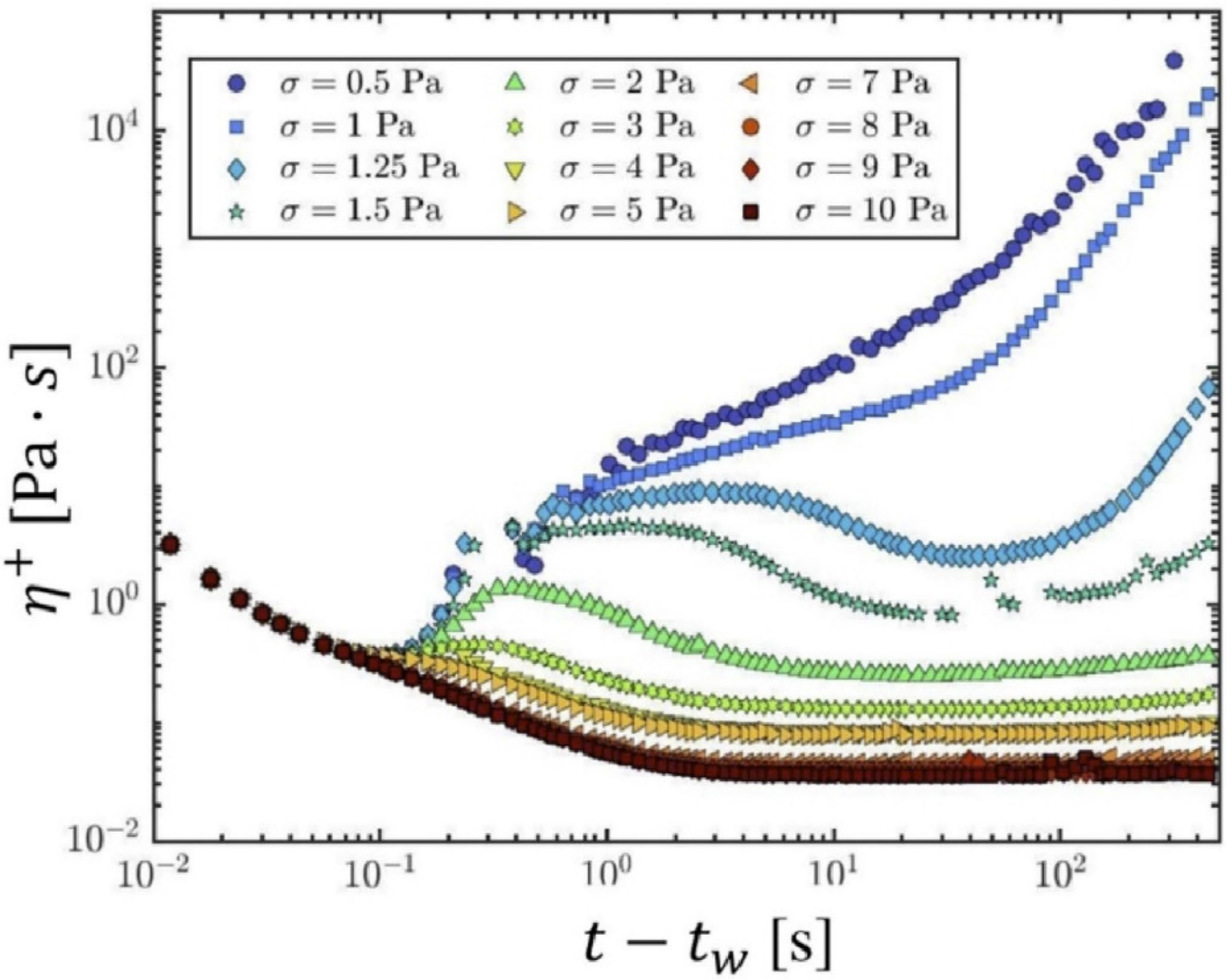

**Figure 7.** Transient viscosity $\eta^+$recorded during creep tests on a 5 wt % aqueous bentonite clay dispersion at various applied stresses, as shown in the legend, as a function of creep time $(t - t_w)$, where $t_w$= 60 s. This figure is adapted, with permission, from Fig. 3(b) of Rathinaraj et al. [75].

Rathinaraj et al. [75] investigated a 5 wt % aqueous dispersion of Wyoming bentonite, a natural clay with plate-shaped montmorillonite particles (about 300–500 nm in size and 1 nm thick). Over time, these particles tend to associate in edge-to-face and edge-to-edge fashion, forming a 'house-of-cards' microstructure that spans the sample [76]. To obtain a limiting structureless state and a uniform initial condition, the sample was first "shear-melted" at a high shear rate, then left undisturbed at 25 °C for a fixed wait time of $t_w$ =60 s. A constant stress ranging from 0.5 to 10 Pa was then applied, and the resulting shear rate was continuously recorded for several hundred seconds to calculate a transient viscosity $\eta^+(t) = \sigma_0/\dot{\gamma}(t - t_w)$.

In Fig. 7, we plot this transient viscosity against creep time on double-logarithmic axes. At the three lowest stresses shown (up to 1.25 Pa), beyond a certain creep time, $\eta^+$ increases continuously after several hundred seconds, and if extrapolated far enough, the viscosity on these branches would formally diverge. This behavior indicates a material whose structure continues to develop while its flow gradually diminishes. At stresses of 3 Pa and higher, $\eta^+$ initially decreases from its starting value and eventually reaches a steady-state plateau representing the flowing, rejuvenated branch. At intermediate stresses such as 1.5 and 2 Pa, the response is more subtle. Shear rate lingers before clearly indicating which branch it belongs to, reflecting a sample near the critical stress that divides the two regimes. Rathinaraj et al. [75] note that for this protocol, the viscosity bifurcation occurs within a critical-stress window of about 1–2 Pa, which varies with the duration of stress application ($t_w$). Since the shear rate on the flowing branch reaches several tens of inverse seconds over tens of seconds, the accumulated strain before reaching steady state surpasses at least 100 strain units, likely reaching into the thousands at higher stresses, which is far beyond the typical strain that can be attributed to viscoelasticity. Figure 7, therefore, illustrates both key features of this stress-controlled form of Definition 1: a clear bifurcation between arresting and flowing states, and the extensive strain needed for the flowing branch to reach a steady state.

Kumar and Joshi [77] studied a chemically distinct system: an aqueous dispersion of Ludox TM-40, a colloidal silica sol with spherical particles of around 20 nm diameter, combined with a 10 wt% aqueous NaCl solution at a 6:13 weight ratio (silica stock to salt solution). The overall composition leads to a silica volume fraction of about 6% and an ionic strength of roughly 1.4 M. Dissociation of salt causes screening of the charges on the particle surface, facilitating the transition from sol to gel within minutes of mixing. The gel was then stored quiescently and studied 1 day after preparation. Kumar and Joshi [77] report that the aging of this gel is reversible by shear melting over storage times of hours but becomes progressively irreversible over days. At a fixed storage time, however, repeated shear melting restores the same initial state as required by our definitions. Before each creep test, the sample was shear-melted at 500/s for 1000 s and subsequently kept under stationary conditions for a fixed waiting time of $t_w$ = 10 s. Afterward, a constant stress (between 0.5–8 Pa) was applied for 1000 s, and the resultant shear rate $\dot{\gamma}$ was recorded.

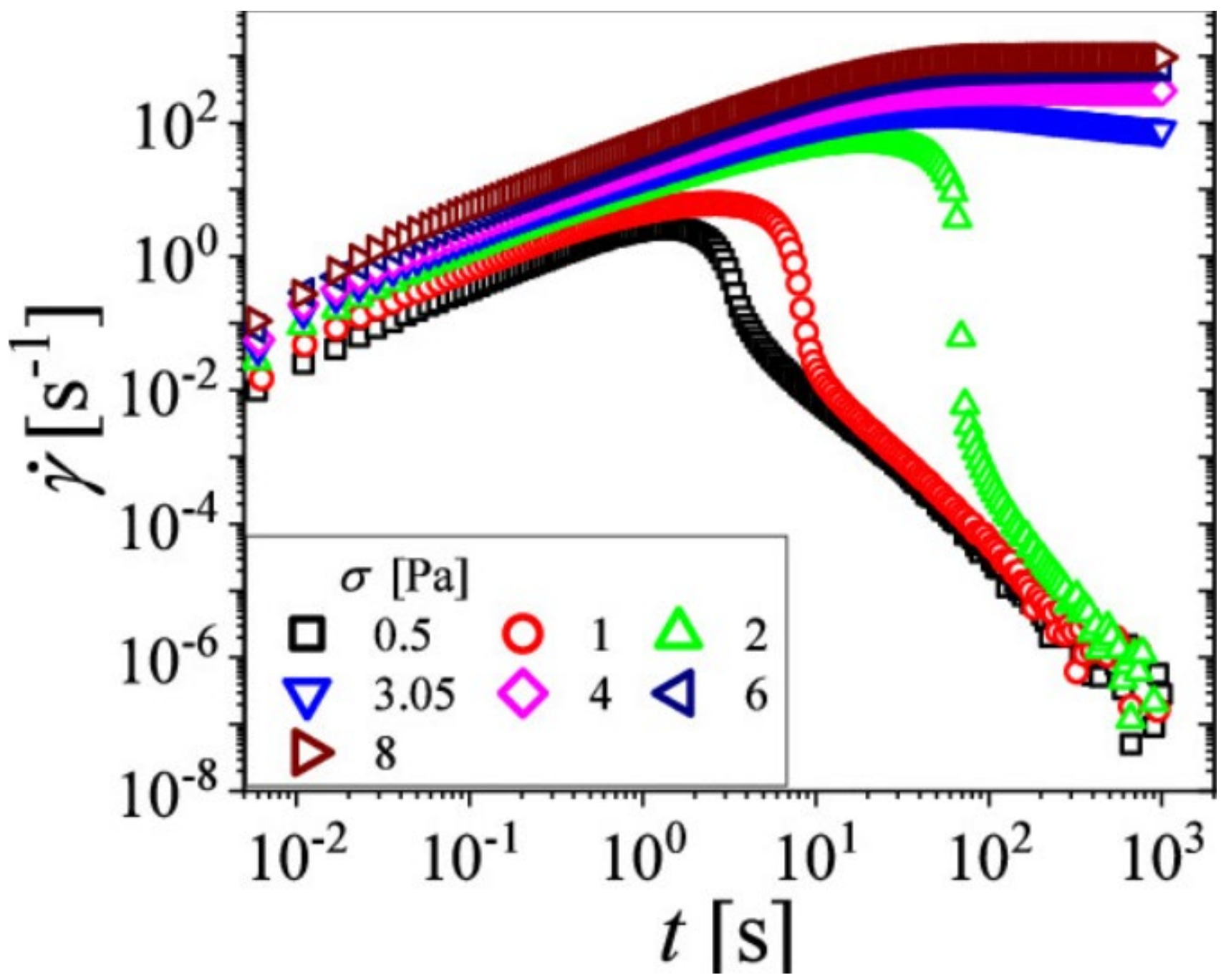

**Figure 8.** Transient shear rate $\dot{\gamma}$ is plotted as a function of creep time under various applied shear stresses, as shown in the legend, for a 1-day-old aqueous silica dispersion (Ludox TM-40). Subsequent to shear melting, the system was kept under quiescent

conditions for 10 s before being subjected to the aforementioned stress. Adapted, with permission, from Fig. 9 of Kumar and Joshi [77].

Fig. 8 illustrates the observed shear-rate transients. For applied stresses up to 2 Pa, the shear rate initially increases, reaches a broad maximum, and then sharply decreases within 1000 s, indicating a brief creep followed by a flow arrest. In contrast, at stresses of 4 Pa and higher, the shear rate increases continuously and attains a steady state, representing the flowing branch. The behavior at 3.05 Pa is particularly illustrative: it closely follows the flowing branch for nearly a minute, with shear rates exceeding $10^2$/s, before eventually decreasing and approaching a possible arrest at longer times. This highlights how close this stress is to the critical value $\sigma_c$ = 4 Pa and how easily a near-threshold system might be mistaken for steady-flow based on limited data. Since the plateau shear rates on the flowing branch are very high, of the order of hundreds of inverse seconds, and sustained for tens of seconds before stabilization, the accumulated strain exceeds thousands of units. This again far exceeds what any typical viscoelastic material will demonstrate. Kumar and Joshi [77] also observed that this critical stress increases with the waiting time before loading, reflecting the same rest-time dependence of the threshold stress noted by Coussot et al [41] for bentonite.

As a further, and final, example of this stress-controlled form of Definition 1, we consider an aqueous multilamellar-vesicle (MLV) suspension, comprised of pH regulated mixture of de-ionized water, surfactant Miracare SLB 365 W, and sodium chloride, studied by Burgeson and Rogers [78]. MLVs form a base for cosmetic products such as shampoos and cleansers [79]. Under shear, this system self-assembles into densely packed multilamellar vesicles that form a weak viscoelastic solid at rest. Burgeson and Rogers [78] subjected the material to creep and obtained not just the total strain response but the recoverable (elastic) and unrecoverable (viscous) components of the same. This was based on the recovery-rheology framework developed by Rogers and coworkers for decomposing transient rheological responses into recoverable and unrecoverable contributions [80, 81]. This decomposition, therefore, provides a direct measurement of the elastic reference strain that Definitions 1 requires. In Figs. 9(a)–(c) the resulting recoverable and unrecoverable strains are plotted as a function of elapsed creep time, for constant applied stresses of 8, 16, and 25 Pa, respectively. It can be seen that at 8 Pa, almost all of the imposed strain remains recoverable throughout the accessible experimental window (1000 s). The corresponding viscous strain can be seen to remain roughly an order of magnitude lower than the elastic strain over this time period. At 16 and 25 Pa, on the other hand, the elastic strain rapidly saturates to a low stress-independent plateau, which also leads to the corresponding recoverable creep modulus, $G_{rec} \approx 105$ Pa across this stress range. Over the explored creep time, the unrecoverable strain continues to grow, at first gradually, but eventually exceeds the recoverable strain by two to five orders of magnitude. Interestingly, to distinguish thixotropic behavior from ordinary nonlinear viscoelasticity, Definition 1 precisely requires a large excess of total strain over the corresponding elastic reference, before steady state is reached at the fixed applied stress. For this material, it is evident that until hundreds or thousands of strain units have accumulated, before the strain rate settles into a constant, steady-state value, if indeed it occurs at all within the experimental window.

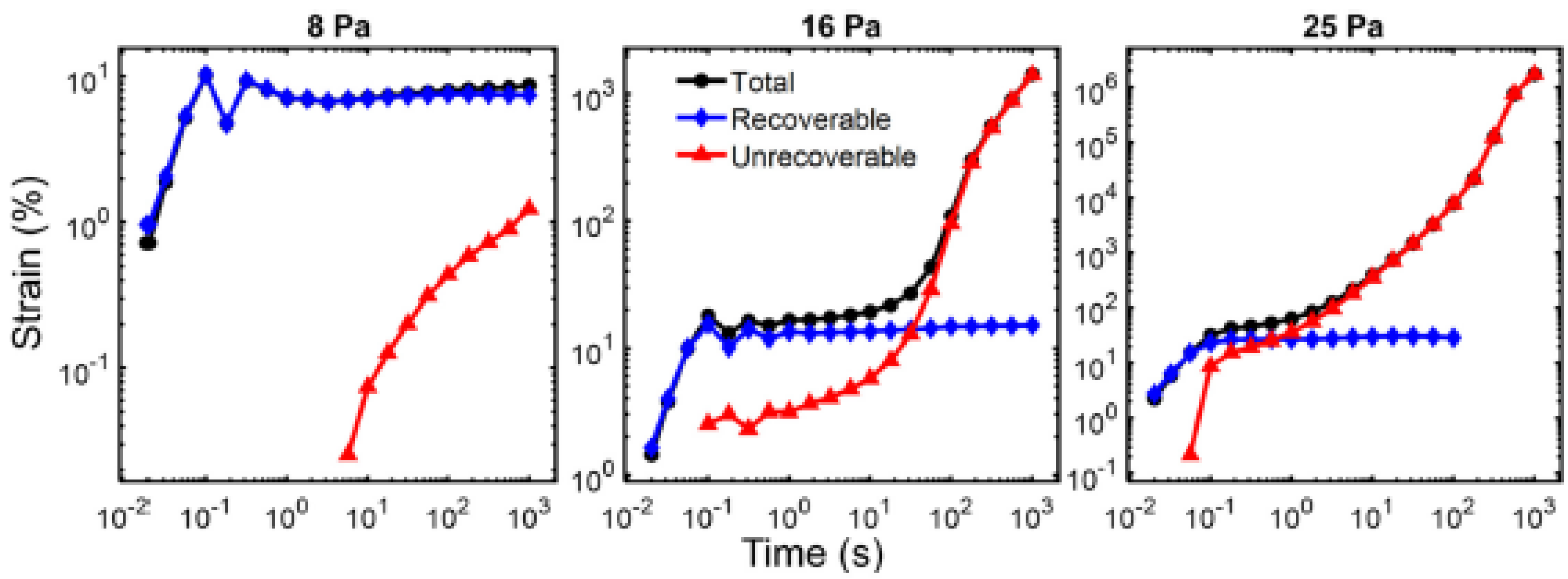

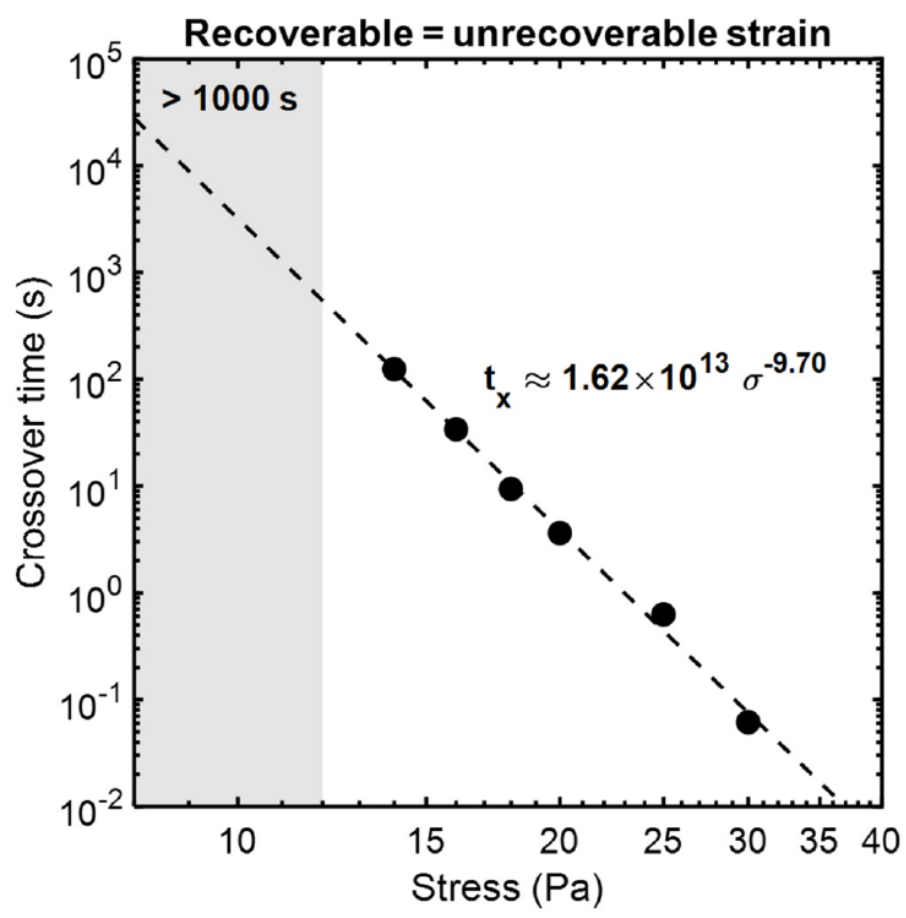

**Figure 9.** Recoverable and unrecoverable strain decomposition for an aqueous multilamellar-vesicle (MLV) surfactant system subjected to constant applied stresses of (a) 8 Pa, (b) 16 Pa, and (c) 25 Pa, obtained from iterative creep-recovery tests. At each elapsed time, the total strain (black) is decomposed into a recoverable, elastic strain (blue) and an unrecoverable, viscous strain (red). (d) Crossover time, $t_x$, at which the recoverable and unrecoverable strains become equal, plotted as a function of applied stress. The dashed line is a power-law fit, $t_x \approx 1.62 \times 10^{13}\ \sigma^{-9.70}$. This figure is adapted, with permission, from Figs. 3 and 4 of Burgeson and Rogers [78].

In Fig. 9(d) the crossover time, $t_x$, defined as the time at which elastic and viscous strain components become equal, is plotted against the applied stress. It can be seen that the crossover time decreases steeply with increasing stress over more than five decades in time. The relationship follows an apparent power law relationship shown by a line: $t_x \approx 1.62 \times 10^{13}\ \sigma^{-9.70}$ for the data points measured beyond 12 Pa. Below that limit, the authors did not observe any crossover within the 1000 s creep window. However, it can be seen that at 8 Pa the unrecoverable strain in Fig. 9(a) continues to increase gradually over the longest times probed. Consequently, these data alone cannot establish whether this stress lies below a true threshold. Discriminating between these two possibilities would require tracking the strain rate itself, rather than the accumulated strain, over a correspondingly longer waiting time. This ambiguity is not a peculiarity of the present dataset but a generic feature of any finite-duration test for a stress threshold. It has indeed been observed that stress-controlled creep responses that appear headed toward one branch can eventually reverse [41, 82, 83]. Therefore, only sufficiently extended observation at even smaller applied stresses can unambiguously resolve whether a truly arrested branch exists within the stress range shown in Fig. 9(d). Hence, we regard the low-stress data as consistent with, but not conclusive proof of, viscosity bifurcation. Importantly, this uncertainty does not affect the classification of the MLV system as thixotropic as it affects only which aspect of our Definition 1 the 8 Pa response most directly corresponds to. If this stress lies on the flowing branch, however slowly the material ultimately flows, it satisfies Definition 1 directly. This is because even this conservative reading already requires a strain of many hundreds of units to reach that state. If it instead lies just below the true threshold, the viscosity and relaxation time diverging there without ever reaching a steady state are precisely the phenomenon Definition 2 addresses directly, which has been discussed below. On the other hand, it can be seen that above about 30 Pa the transition to unrecoverable flow becomes nearly instantaneous. To summarize, in this example (i) the elastic reference strain is obtained from the same protocol as the thixotropic response itself. Furthermore, (ii) at higher stresses, the unrecoverable strain exceeds the elastic strain by orders of magnitude before a steady flow is established, if one is established at all within the experimental window. In addition (iii) the stress-dependence of the crossover time in Fig. 9(d) is consistent with a viscosity bifurcation, although, as noted above, the low-stress data do not conclusively establish one. Considering these aspects, we believe this example of MLV is a clear realization of stress-controlled form of Definition 1. Moreover, with regard to its lowest-stress behavior, should it prove to be a true arrest rather than an extremely slow flow, it would also constitute a realization of Definition 2.

Figs. 7–9 collectively demonstrate the stress-controlled form of Definition 1 across three chemically and structurally distinct systems: a natural, plate-shaped aluminosilicate clay, a synthetic, roughly spherical silica nanoparticle gel, and a self-assembled, lamellar surfactant network leading to MLV. In all three cases, a threshold stress divides an aging branch, where the unrecoverable strain grows without an apparent bound, from a rejuvenated, flowing branch, where the material only reaches a steady shearing state after strains ranging from several hundred to several thousand units. On the flowing branch, in each case, the observed strain threshold for steady state is significantly beyond what viscoelasticity alone can explain, exactly the criterion specified by Definition 1, now based on a controlled stress rather than a controlled rate. Notably, unlike the clay and silica examples, the multilamellar vesicle system additionally allows the elastic reference strain to be measured directly by performing strain recovery experiments, rather than being estimated from a separate measurement, offering a particularly rigorous demonstration of the criterion. The threshold stress itself is not a static material constant; it increases with the aging or waiting time allowed before the stress is applied, consistent with the waiting-time sensitivity reported for bentonite suspension by Coussot and coworkers [41]. Taken together with the six rate-controlled

examples above, the three stress-controlled examples show that large-strain criterion of Definition 1 holds regardless of which variable, shear rate or stress, is used to control the deformation. One aspect of the phenomenon, however, falls outside the purview of Definition 1 in the strict sense and that is the aging branch, on which viscosity and relaxation time grow without bound and no steady state is ever reached. Since no steady state exists to be reached on the aging branch it is not a case of observation of large strain. Instead, it leads to the direct experimental evidence for Definition 2, which we discuss in the next subsection.

**III.B. Definition 2**

Definition 1 requires imposing a specific, sufficiently large, deformation, whether a step-up in shear rate or a step change in stress, and then monitoring the evolution of viscosity toward a new steady value, or, on the aging branch of its stress-controlled form, its divergence instead. If a steady state is reached, its attainment must occur under an anomalously large strain, relative to what can be attributed to viscoelasticity. Interestingly, Definition 2 removes this strain requirement entirely. Instead of asking how much deformation is needed to erase the memory of a fixed reference state, it considers a simpler, more intrinsic question: does the material's own relaxation time $\tau$, the timescale over which the material retains memory of a prior shear or stress history, increase without bound as the quiescent waiting time $t_w$ elapsed since that history was imposed keeps increasing? According to Definition 2, this growth in $\tau$ continues regardless of how large $t_w$ becomes, as revealed by rheological measurements performed after the rest period. Because $\tau$ cannot be probed directly in the way dynamic moduli can, its unbounded growth must be inferred from a response function, the relaxation modulus ($G$) or the creep compliance ($J$), that depends explicitly on $t_w$ rather than only on $t - t_w$, leading to dependencies such as $G(t - t_w, t_w)$ or $J(t - t_w, t_w)$; more generally, it can be any measurable quantity that is known to correlate with rheology, including a scattering correlation function, whose characteristic decay time depends explicitly on $t_w$. Provided the underlying microstructural rearrangement is itself at least partially reversible under shear, a requirement common to all three definitions and discussed at the start of section III, such a material is thixotropic according to Definition 2. This criterion applies to both stress- and strain-controlled protocols. Non-rheological protocols, such as dynamic light scattering, diffusing-wave spectroscopy, or microrheology, can also be admitted as evidence for Definition 2, but only to the extent that the structural or optical relaxation time they report has been shown, or can reasonably be argued, to correlate with the material's mechanical relaxation time. The aging branches already shown in Figs. 7–9, on which the viscosity keeps growing without reaching a steady state, are consistent with this criterion, since continued growth of viscosity under constant stress also implies increase in the relaxation time. This is, however, indirect evidence as in order to obtain of evolution of relaxation time, a response function is needed to be measured at several waiting times, as done in the examples mentioned below.

Any material relaxes after being subjected to a perturbation, and probing it again before this relaxation is complete will always trivially show a $t_w$ dependence. Even if two small strains are imposed on a non-thixotropic linear viscoelastic material with a waiting time $\mathrm{t_w}$ separating them, the effect of $\mathrm{t_w}$ will be manifest in the Boltzmann linear superposition of the responses to each strain. However, once $\mathrm{t_w}$ surpasses the longest relaxation time $\tau_\mathrm{r}$ of the material, the response of such a material depends only on $\mathrm{t} - \mathrm{t_w}$. In contrast, a thixotropic material shows an additional dependence on $\mathrm{t_w}$, often because its relaxation time grows with $\mathrm{t_w}$. Various examples shown in this subsection that validate Definition 2 are investigated over significant time durations, sometimes over several hours of $\mathrm{t_w}$, and in every case, the $\mathrm{t_w}$-dependence has been observed to show no signs of saturating.

A further clarification is necessary regarding how this waiting-time dependence is best analyzed. In the following discussion, we present several of the examples that track the storage and loss moduli ($G'$ and $G''$) at a fixed frequency, or a yield stress ($\sigma_y$) extracted from a fixed-frequency amplitude sweep, as a function of $t_w$. These properties are measured because they are convenient probes, but, strictly speaking, they are indirect proxies for the time dependent thixotropic structural evolution. The property fundamentally responsible for thixotropic behavior is the material's characteristic relaxation time $\tau$, which is reflected in its zero-shear viscosity $\eta_0 \sim G_0\tau$. The continued unbounded growth of relaxation time and/or viscosity with waiting time and shear reversibility of the same indicate the existence of thixotropy. A modulus measured at a single, fixed frequency $\omega$ is sensitive chiefly to the elastic strength $G_0$ of the microstructure over the associated timescale $\approx 1/\omega$. Once $t_w$ is large enough that $\omega\tau(t_w) \gg 1$, the material already appears fully elastic at that frequency. Consequently, $G'(\omega, t_w)$ probed under practical values of $\omega$ can plateau even while $\tau(t_w)$ itself, when measured more directly through creep or stress-relaxation experiments repeated at different $t_w$, continues to grow with no indication of a limit. On the other hand, some standard physically aging systems such as hard sphere dispersion may not show any distinct increase in dynamic moduli or of yield stress as a function of waiting time. Therefore, an apparent saturation of $G'$, $G''$, or $\sigma_y$ at long waiting times should not, therefore, be taken on its own as evidence against thixotropy. The more rigorous test, hence, involves obtaining time dependence of relaxation time itself as obtained by various groups in the literature as noted in the recent review [55]. However, no experiment can run for an unbounded time. What can be established in practice is only the absence of saturation over the accessible experimental window, several decades of $t_w$ in the examples that follow, which we take as strong, operational evidence for the underlying, unbounded growth that Definition 2 proposes.

At this point it is important to note an unavoidable limitation of this operational criterion. Let us consider a material, whose true equilibrium relaxation time is extremely large but nonetheless finite. Now, if this equilibrium value exceeds the maximum waiting time accessible in a laboratory experiment, the measured relaxation time would necessarily appear to grow without any sign of saturation over the entire accessible window. This situation, therefore, mimics a genuine unbounded growth even though the material is not, strictly speaking, thixotropic by Definition 2. No finite-duration experiment can, even in principle, distinguish this scenario from true divergence, since both predict identical behavior over any bounded observation time. Note that, this limitation is not specific to Definition 2. It is a generic feature of any operational test for a diverging physical quantity. For instance, there is a long-standing debate over whether a "true" yield stress exists or whether all apparently yield-stress materials instead possess an extremely large, but finite, low-shear-rate viscosity [84]. We, therefore, do not claim that absence of saturation over an experimentally accessible window constitutes proof of literal divergence. We, instead, regard it as the strongest evidence practically obtainable, and adopt it as the operational standard for Definition 2 throughout this work. We recommend that researchers explicitly report the maximum waiting time explored, ideally spanning as many decades of $t_w$ as feasible, so that readers can judge the strength of the operational evidence for themselves. We further note that, for most practical purposes, such as material processing, storage, and shelf life, the distinction between a truly diverging relaxation time and one that is merely very large relative to all timescales of use may be of limited practical consequence. Examples wherein waiting time dependence of the relaxation time has been obtained explicitly offer the most complete demonstration of Definition 2. On the other hand, where only a modulus or yield stress is reported, the observed growth remains meaningful, but is best read as a working proxy rather than a direct measurement of the underlying, diverging relaxation time.

We begin with five examples that share a common protocol. In each case, the material is aged for a waiting time $t_w$ after cessation of shear, and then probed by either a step stress or a step strain. The resulting creep compliance or relaxation modulus is then examined as a function of both $t_w$ and the time elapsed since the probe was applied $(t - t_w)$. In each of these cases the authors analyze the response function and extract a power-law growth of the underlying relaxation time: $\tau \sim t_w^{\mu}$, leading to the most direct, quantitative evidence for Definition 2.

The first of these examples is a near-hard-sphere colloidal suspension studied by Derec et al. [85]. They prepared a 41.7 vol. % aqueous dispersion of aqueous suspension of silica particles (≈100 nm diameter) stabilized by adsorbed poly(ethylene oxide). Silica volume fraction of 41.7% in the dispersion is above the critical volume fraction of about 40.5% beyond which the suspension develops a yield stress and undergoes physical aging. The sample was mechanically rejuvenated by a large oscillatory shear deformation (strain amplitude of 1000% at frequency of 1 Hz) for 100 s. Subsequently, it was left to rest under stationary conditions in the rheometer. After a waiting time $t_{\mathrm{w}}$ it was subjected to step strain of 1% within the linear regime, and the following relaxation of the stress was recorded as a function of the time t' elapsed since application of the step strain.

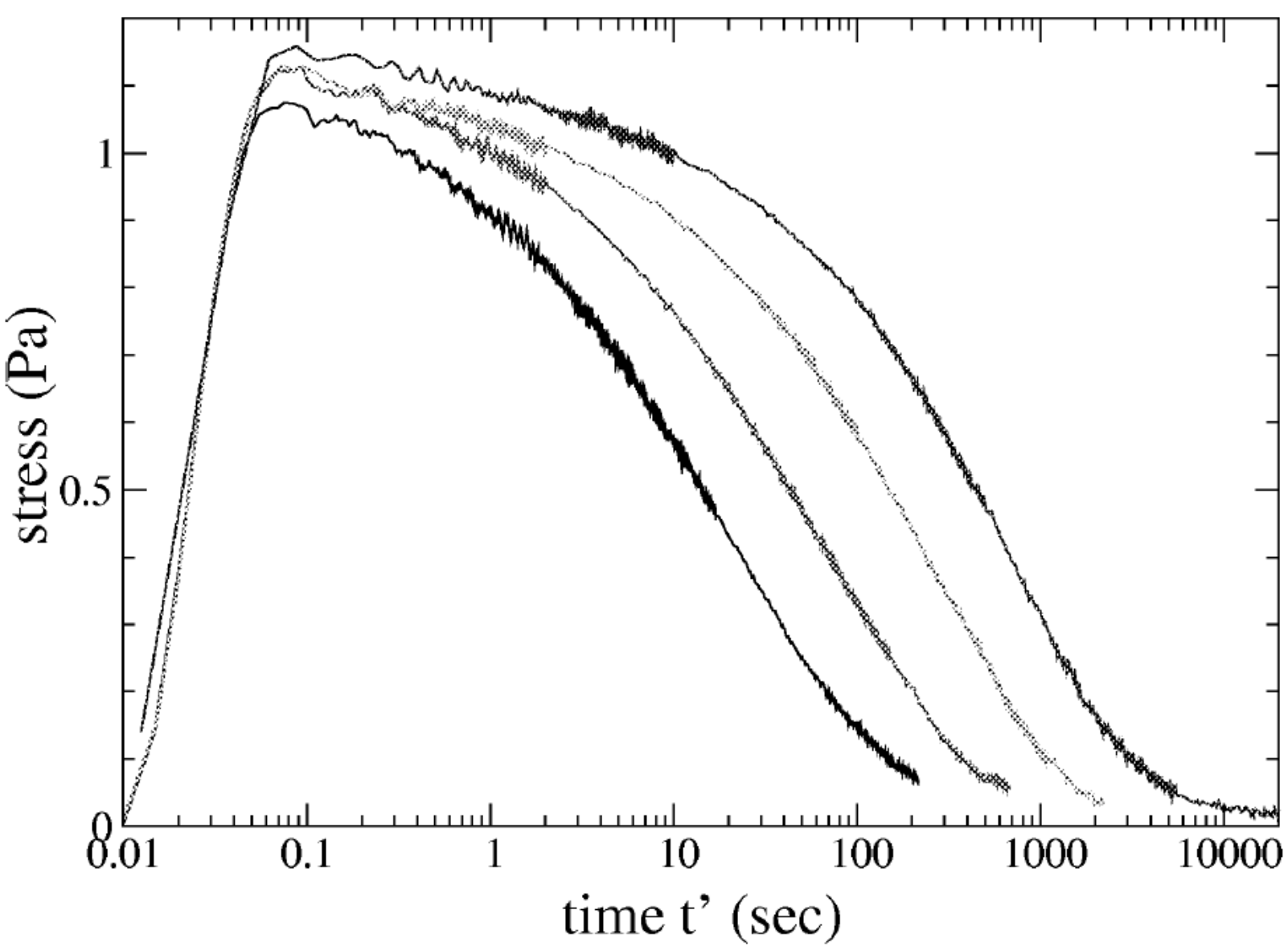

**Figure 10.** Stress relaxation $\sigma(t')$ following a step strain of 1%, applied at four different waiting times $t_w$ = 10, 100, 1000, and 10,000 s (from left to right) after mechanical rejuvenation (large-amplitude oscillatory shear at 1000%, 1 Hz, for 100 s), for a silica suspension (41.7 vol. %) stabilized by adsorbed poly(ethylene oxide). Adapted, with permission, from Fig. 3 of Derec et al. [85].

Fig. 10 shows that after application of the step strain, a brief instrumental inertial transient lasts for less than 0.1 s beyond which the stress attains the highest value. Subsequently, the stress relaxes, albeit more slowly for the samples that have been left to rest for a longer duration. It can be seen that four curves for $t_w$ spanning three orders of magnitude are systematically shifted to later times without changing their qualitative shape. Interestingly, Derec et al. [85] further show that all four relaxation curves collapse onto a single master curve upon appropriate normalization. This analysis suggests that the relaxation time of this system shows a power-law dependence on waiting time, with an exponent of 0.55. This shows that a single, small step-strain relaxation test, repeated at different waiting times, is sufficient to show that the relaxation modulus $G(t';\, t_w)$ depends explicitly, and without any bound, on the waiting time $t_w$ itself. This is exactly the criterion that Definition 2 requires, and it is worth emphasizing that this is precisely the direct measurement of relaxation-time growth called for at the beginning of this section.

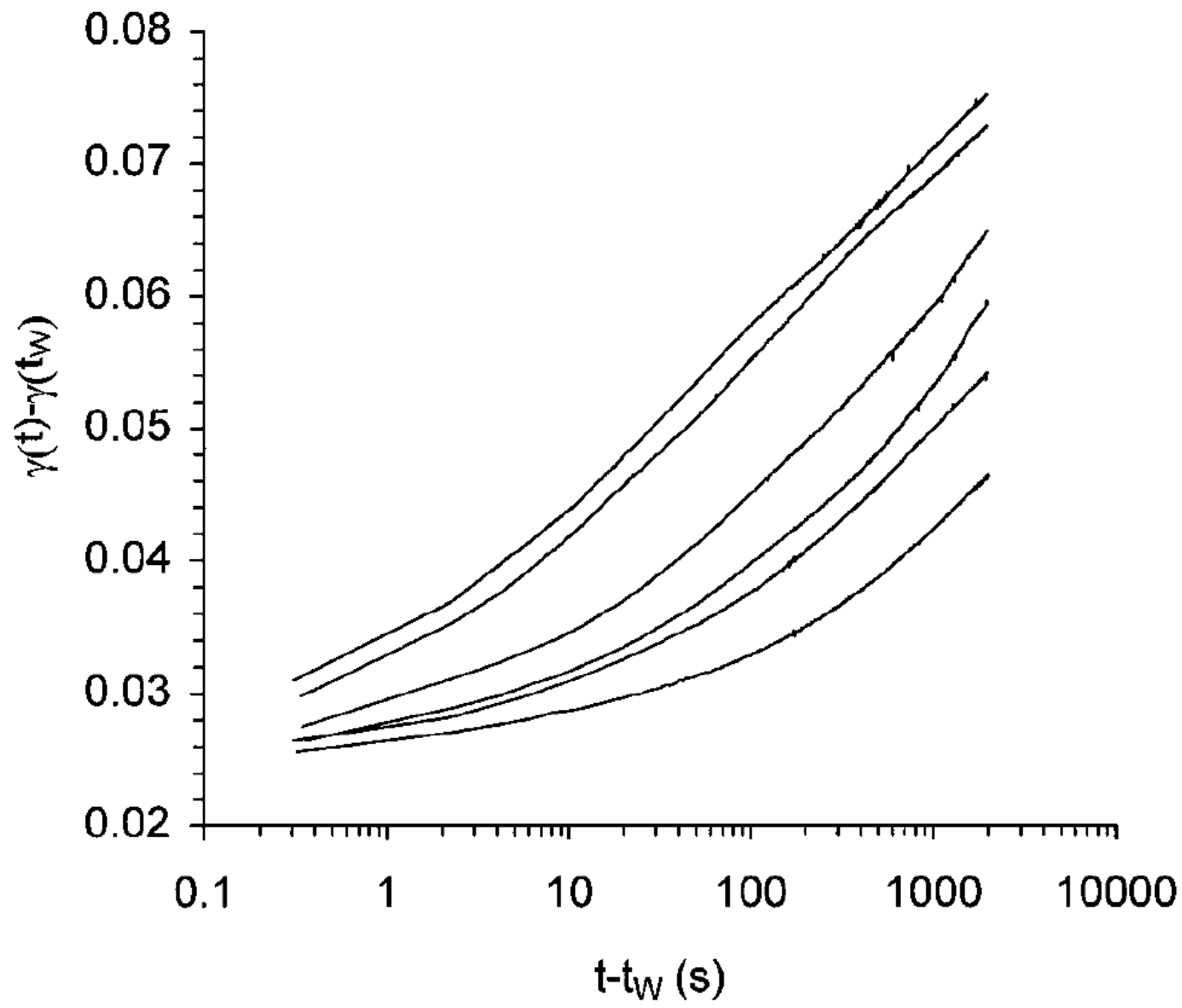

**Figure 11.** Creep response $\gamma(t) - \gamma(t_{\mathrm{w}})$ upon application of the stress of 10 Pa applied at waiting times $t_{\mathrm{w}}$ = 15, 30, 300, 1000, 2000, and 10,000 s (top to bottom) following cessation of mechanical rejuvenation at a stress exceeding the yield stress applied to a paste of cross-linked acrylate microgel particles. Adapted, with permission, from Fig. 3 of Cloitre et al. [86].

In the second example, concentrated pastes of soft, deformable polyelectrolyte microgels, consisting of cross-linked acrylate particles, that swell in water were probed by creep by Cloitre et al. [86]. When these particles are packed beyond their overlap concentration, they de-swell and develop flat facets, squeezing neighboring particles to produce a disordered, jammed paste with a well-defined yield stress $\sigma_{\mathrm{y}} = 20$ Pa. These microgel samples were first mechanically rejuvenated by applying a stress exceeding $\sigma_{\mathrm{y}}$ for 60 s. After shear melting, the microgel paste was left undisturbed in the rheometer. At the waiting time $t_{\mathrm{w}}$ subsequent to cessation of shear melting, a constant stress, smaller than $\sigma_{\mathrm{y}}$was applied, and the resulting strain $\left(\gamma(t) - \gamma(t_{\mathrm{w}})\right)$ beyond that at $t_{\mathrm{w}}$ was measured over time. Fig. 11 shows two distinct signatures of waiting-time dependence. Firstly, the instantaneous strain upon application of the probe stress, which is essentially elastic, is largest for the shortest $t_{\mathrm{w}}$ and decreases with increasing $t_{\mathrm{w}}$. This behavior reflects the time-dependent increase in the material's elastic modulus. Second, and more importantly, the rate of increase in strain decreases with increasing preceding waiting time. Cloitre et al. report that this probe stress (10 Pa) lies between a lower bound $\sigma_c \approx 4$ Pa, below which the response shows undisturbed aging (an exponent of $\mu \approx 1$), and the yield stress $\sigma_y$ = 20 Pa, above which aging is fully erased (exponent $\mu = 0$). The value of $\mu = 0.8$ reported for the stress of 10 Pa, therefore, reflects partial rejuvenation by the probe stress itself, which is a direct demonstration of the reversibility that Definition 2 requires. It can be seen that the curves for $t_\mathrm{w}$ spanning nearly three decades remain clearly separated throughout the accessible window, with no indication that they are converging toward a common, $t_\mathrm{w}$-independent response. Cloitre et al. [86] show that these creep curves collapse onto a single master curve when creep time is normalized by a function that can be interpreted as the time-dependent relaxation time of a material. Interestingly, for the data shown in Fig. 11, this increase in relaxation time follows a power-law in $t_\mathrm{w}$, with an exponent of 0.8. We believe that this is a further, and unusually explicit, realization of Definition 2. In this case, not only does the strain response depend on waiting time without bound, but the corresponding relaxation time has been reported to show power law dependence on waiting time.

In the third example, both the protocols: creep as well as stress relaxation were performed on a single material. Kaushal and Joshi [87] studied a commercial concentrated acrylic polymer-in-water emulsion paint (Kansai Nerolac Paints Limited, Mumbai; 48% w/w solid content), whose physical aging likely arises from network formation by surfactant-stabilized acrylic particles suspended in aqueous media. The sample was shear-melted with a large-amplitude oscillatory strain (30000 % at a frequency of 0.1 Hz) to break the structure and erase the prior deformation history. After a predetermined waiting time, $t_w$ of up to 90 min, they subjected the material to either a step stress of 2 Pa or a small step strain of 1% and the corresponding creep compliance $J(t - t_w)$ and stress-relaxation modulus $G(t - t_w)$ were recorded at different values of $t_w$.

Fig. 12 plots both response functions against elapsed time $t - t_w$ (since application of either step strain or stress) after different waiting times $t_w$, as shown in the legend. It can be seen that, notwithstanding the waiting time, the compliance increases while the relaxation modulus decreases with $t - t_w$, as expected of any viscoelastic material. On the other hand, at any given $t - t_w$, the compliance shifts to lower values while the relaxation modulus shifts to higher values systematically with increase in waiting time $t_w$ without showing any noticeable sign of saturation. This behavior is attributed to physical aging, which results in progressively slower relaxation dynamics and a higher storage modulus as the waiting time increases. Because $J$ and $G$ obtained in this work are explicit functions of both $t - t_w$ and $t_w$, leading to $J = J(t - t_w, t_w)$ and $G = G(t - t_w, t_w)$ and because this dependence is tracked well beyond any timescale one could reasonably attribute to ordinary viscoelastic relaxation, this system satisfies the criterion proposed by Definition 2.

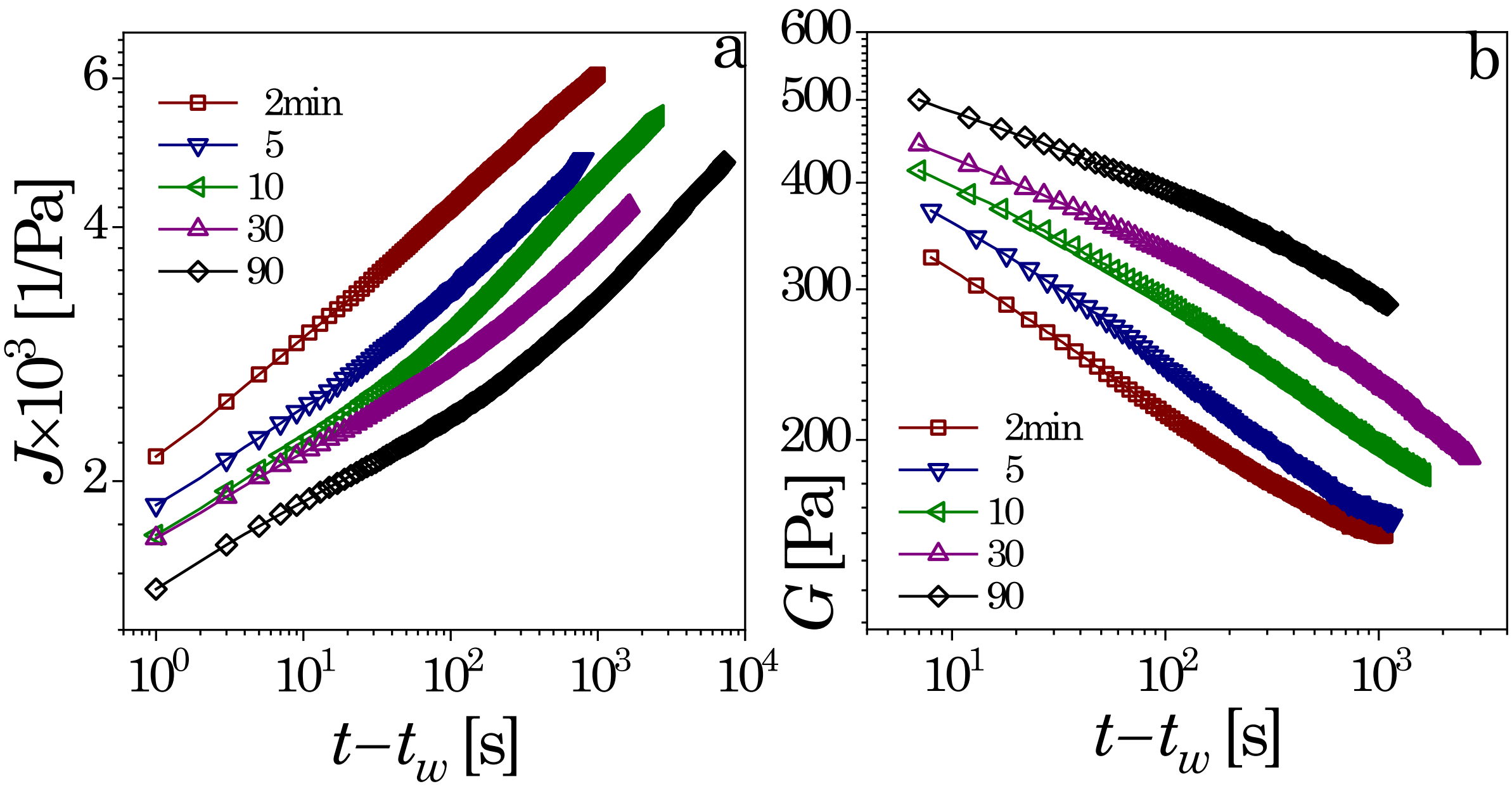

**Figure 12**. Waiting-time dependence of (a) creep compliance at an applied stress of 2 Pa and (b) stress-relaxation modulus at an applied strain of 1%, plotted against time elapsed since application of the step stress or step strain, for an acrylic emulsion paint (48 mass % dispersed-phase content) shear-melted by a large-amplitude oscillatory strain, at the waiting times shown in the legend. Adapted, with permission, from the Supplementary Information (Fig. S1) of Kaushal and Joshi [87].

It is important to note that Kaushal and Joshi [87] transformed both the response functions $J = J(t - t_w, t_w)$ and $G = G(t - t_w, t_w)$, which violate the Boltzmann superposition principle owing to additional $t_w$ dependence, from real time to *effective time* $\xi(t)$ defined as: $\xi(t) = \tau_0 \int_0^t \frac{ds}{\tau(s)}$, where $\tau(s)$ is the real time-dependent relaxation time while $\tau_0$ is the constant relaxation time associated with the effective time. Kaushal and Joshi [87] observed that this framework validates the Boltzmann superposition principle in the effective time domain and in this domain leads to time-waiting time superposition when a vertically shifted relaxation modulus or compliance is plotted as a function of effective time elapsed since application of the deformation field ($\xi(t) - \xi(t_w)$). Through this analysis, they find that the relaxation time scales as waiting time by: $\tau \sim t_w^{1.19}$, a direct, quantitative confirmation of the unbounded relaxation-time growth that is the essential signature of Definition 2. Interestingly, the power law exponent exceeds unity ($\tau$ grows even faster than linearly with $t_w$), and the material is therefore classified as "hyper-aging." Remarkably, they report that the convolution relationship is valid in the effective time domain, enabling the prediction of compliance from relaxation modulus and vice versa. This example suggests that this material preserves the shape of the relaxation-time distribution as it undergoes thixotropic restructuring over time. This feature, while not a necessary requirement of thixotropy, has been reported for many materials.

A fourth and fifth example studies behavior of two inorganic colloidal gels. Kumar et al. [88] studied two soft colloidal gels, a 30-day-old, 3.5 wt % aqueous Laponite clay dispersion, and a 2-day-old, 6 vol % Ludox TM-40 silica gel (silica stock combined with 10 wt % saline water in a 6:13 weight ratio). Following thorough shear melting of each sample (100/s for 1000 s for the clay dispersion, 1000/s for 1000 s for the silica gel), the authors let the samples rest under quiescent conditions for a waiting time $t_w$ and then subjected the same to a constant shear stress well below the material's yield stress (1 Pa for the clay dispersion and 0.5 Pa for the silica gel). Subsequently, they record the resulting creep compliance $J(t - t_w)$ as a function of the elapsed creep time at different $t_w$. For this gel too, shear melting restores a reproducible initial state at fixed age, although the aging becomes irreversible on longer time scales.

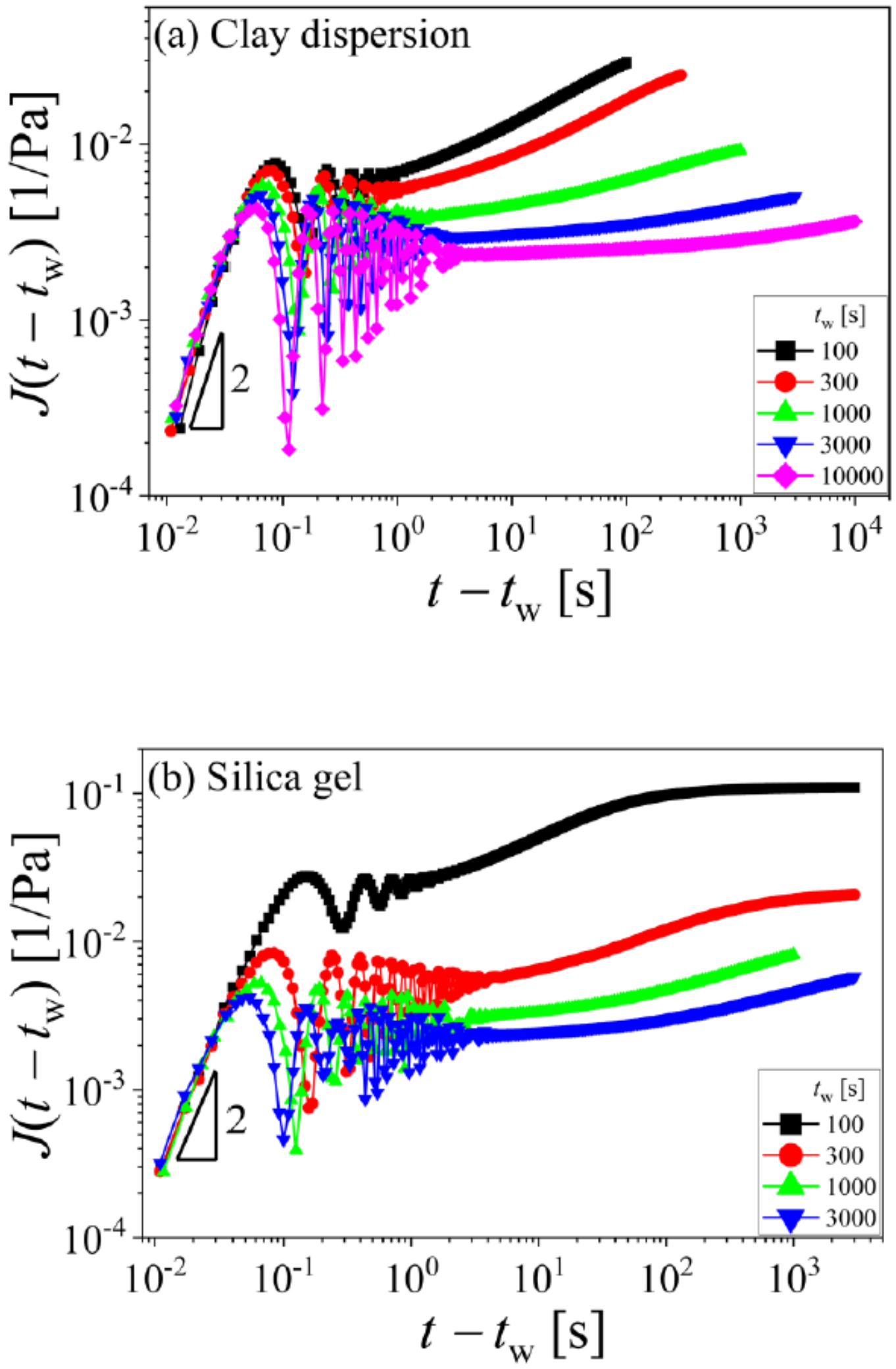

**Figure 13.** Evolution of creep compliance $J(t - t_w)$ with creep time $(t - t_w)$ at different waiting times $t_w$ for (a) a 3.5 wt % aqueous Laponite dispersion (clay dispersion) at an applied stress of 1 Pa, and (b) a 6 vol % Ludox TM-40 silica gel at an applied stress of 0.5 Pa. Legends show the values of $t_w$ at which the creep experiments were performed. Adapted, with permission, from Fig. 6 of Kumar et al. [88].

In Fig. 13, we plot the creep compliance for both materials as a function of creep time $t - t_w$ at different $t_w$. In this figure the compliance demonstrates creep ringing at timescales below O(1) s, preceded by a quadratic compliance response at O(0.01) s, arising from a coupling between sample elasticity–instrument inertia associated with the stress controlled rheometer [89, 90]. It can be seen that compliance decreases systematically with increase in $t_w$ when compared at constant creep time. In addition, the rate at which compliance increases with $t - t_w$ also weakens with increasing $t_w$. Since $J$ is a function of both $t - t_w$ and $t_w$, rather than of $t - t_w$ alone, leading to $J =$ $J(t - t_w, t_w)$, the response violates time-translational invariance and hence violates the Boltzmann superposition principle in the real time domain. Kumar et al. [88] further show that this additional waiting-time dependence vanishes once the reference frame is transformed from the real time to the effective time domain. The resulting master curves are consistent with a power-law growth: $\tau \sim t_w^{\mu}$, with $\mu = 1.89 \pm 0.05$ for the clay dispersion and $\mu = 1.24 \pm 0.06$ for the silica gel. Similar to the previous example, both materials show hyper-aging behavior. In any case, since relaxation time grows with $t_w$, it never settles to a finite value, and the rheological response of the materials always shows an additional dependence on $t_w$. This is a direct, quantitative confirmation of the unbounded increase in relaxation-time that is central to Definition 2. This example also connects to Definition 1, since Kumar et al. [88] separately report that the same systems show a viscosity-bifurcation, in the stress-controlled sense discussed under Definition 1, at stresses of 42.5 Pa for the clay dispersion (waiting time 100 s) and 2.9 Pa for the silica gel (waiting time 5 s) that are well above the 1 Pa and 0.5 Pa probe stresses used to obtain data shown in Fig. 13. This confirms that the creep measurements in Fig. 13 were performed entirely on the aging, non-flowing branch of the material's response, precisely the branch that Definition 2 addresses.

The trend reported in all five examples above is corroborated by Agarwal et al. [13], who assessed whether shear-thinning viscoelasticity could be mistaken for thixotropy by using an aqueous polyethylene oxide (PEO) solution as an experimental model and finitely extensible nonlinear elastic–Peterlin (FENE-P) constitutive equation [23] as a theoretical model. They independently probed both the systems by step strain and step stress applied after different waiting times and compared the resultant relaxation moduli and creep compliances at a fixed elapsed time $(t - t_w)$. Importantly, the PEO solution and the FENE-P model show only a bounded, monotonic approach to a $t_w$ independent value once $t_w$ exceeds the relaxation time. In contrast, they observed that for a thixotropic Laponite clay dispersion the relaxation modulus and creep compliance shift indefinitely with $t_w$, never settling into a $t_w$-independent response. This is precisely the criterion proposed in Definition 2. We also note that the behavior discussed in this subsection is analogous to the behavior of small-molecule and polymeric glasses, which also show unbounded growth in relaxation time with rest period (but may not show shear rejuvenation), and to the Soft Glassy Rheology (SGR) [61] framework (that shows physical aging as well as shear rejuvenation) discussed in Section II. We, therefore, take SGR framework as further support for Definition 2, extended here as a broader class of thixotropic materials.

Another possible source of confusion in terminology can arise from the observation that "viscoelastic" constitutive equations sometimes include time and stress or strain-rate dependent relaxation times [23], which can be included within differential or integral viscoelastic constitutive equations. Typically, the time dependences of the relaxation time introduced in such equations remains bounded, and hence, by Definition 2, are not "thixotropic." But, if such time dependences were to allow for unbounded growth of the relaxation time, then these constitutive equations would be classified as "thixotropic," under our Definition 2, although they would also be "viscoelastic." Being both viscoelastic and thixotropic should not be considered a contradiction, since virtually all materials considered to be thixotropic also possess properties that are normally considered viscoelastic, such as finite values of $G'$ in linear oscillatory shear. Clarity in terminology will therefore require going beyond an "either-or" understanding of viscoelasticity and thixotropy. The key is to distinguish which aspects of the rheological behavior go beyond viscoelasticity and should be designated "thixotropy." If a nominally viscoelastic constitutive equation contains a shear-rate-dependent relaxation time that diverges at rest, the constitutive model acquires thixotropic, as well as viscoelastic, character. This, and the above discussion by Agarwal et al. [13] show the importance of developing precise definitions of thixotropy, to discriminate when seemingly similar behavior of materials or constitutive equations should be classified differently, and which rheological properties of a material are viscoelastic, and which are thixotropic.

Beyond the rheological measurements discussed above, the relaxation time can also be measured directly through non-rheological, scattering- or tracer-based techniques. Because Definition 2 is framed in terms of the mechanical relaxation time, such probes are admitted as evidence only when the characteristic time they report has been shown, or can reasonably be argued, to correlate closely with the material's rheological relaxation time. Interestingly, an increase in relaxation time and its mechanical reversibility have been documented through the use of non-rheological response functions. In aqueous Laponite clay dispersions, dynamic light scattering has been used to track the intensity autocorrelation function of scattered light following shear-induced rejuvenation (e.g., vigorous shaking or stirring). The characteristic decay time of the intensity autocorrelation function, which is a direct measure of the structural relaxation time, has been observed to grow systematically with waiting time with no sign of saturation over the accessible window [91]. A closely related phenomenon has been reported by van Megen and coworkers [92] for concentrated, nearly hard-sphere colloidal glasses, such as sterically stabilized spherical particles near the glass transition. The authors infer the structural ($\alpha$) relaxation time from the waiting-time-dependent shift of the long-time decay of the self-intermediate scattering function. Within the glassy state, they observe that the characteristic structural relaxation time increases with waiting time after cessation of shear, with no sign of saturation over many weeks [92]. Using multi-speckle diffusing-wave spectroscopy on a different colloidal glass (concentrated polystyrene microspheres stabilized by charge and a steric coating) Lequeux and coworkers [93, 94] directly demonstrated both aspects of the Definition 2. They reported that applying shear resets the growing relaxation time, and once the shear stops, physical aging resumes wherein the relaxation time again grows with the waiting time [93, 94]. A more directly validated bridge between a non-rheological probe and the mechanical relaxation time is provided by passive microrheology. Vyas et al. [95] tracked the thermal motion of embedded tracer particles in aging aqueous Laponite suspensions. They converted the resulting mean-squared displacement into a creep compliance through the generalized Stokes–Einstein relation, and thereafter into a relaxation modulus through the linear-viscoelastic convolution relation. By rescaling time through an age-dependent relaxation time, they showed that the microrheological response at different ages and concentrations collapses onto a single master curve. Through these measurements they dependence of relaxation time on waiting time that shows an unbounded growth. Since this technique converts a tracer measurement directly into creep compliance, it offers a more directly validated estimation of the material's mechanical relaxation time. These representative studies show that, when validated in this way against the material's own rheological response, non-rheological structural and microrheological probes can provide admissible evidence for the unbounded, flow-reversible growth of the mechanical relaxation time central to Definition 2. We note however, that these non-rheological probes must be applied to deformed states of the material to infer that the mechanical rejuvenation needed for a material to be thixotropic, is satisfied.

This subsection has presented various examples that support Definition 2. The material classes that we considered show its applicability across a broad spectrum of materials, including clay and silica colloidal gels, a near hard sphere suspension, a microgel paste, and a commercial acrylic paint, and, via non-rheological scattering and microrheology probes validated against mechanical measurements, Laponite dispersions and hard-sphere colloidal glasses. These materials were examined primarily through application of step-stress or step-strain (creep or stress-relaxation) protocols applied at different waiting times. In all cases, the response function explicitly depends on the waiting time tw, showing no signs of saturation over long waiting times, long enough to take as strong evidence for the unbounded growth of the material's relaxation time that Definition 2 requires. Unlike Definition 1, which mandates a specific deformation protocol, whether rate- or stress-controlled, and a correspondingly large strain, Definition 2 requires only that the material's mechanical relaxation time be probed for long enough, and that the thixotropic aging be reversible by subsequent nonlinear deformation.

Overall, Section III has examined several representative experimental systems and organized them under the two operational, semi-quantitative, definitions of thixotropy proposed in Section II. Under Definition 1, in its rate-controlled form, the analyzed systems were shown to require a significant strain to reach a steady state that exceeded a viscoelastic reference strain by two to five orders of magnitude in every case. Under Definition 1's stress-controlled form, three further, chemically and structurally distinct systems, two colloidal gels and a self-assembled surfactant network resulting in MLV, were each shown to bifurcate into an aging, structurally arresting branch and a rejuvenated, flowing branch, with the flowing branch, in each case, reached only over a strain much larger than the elastic strain. In the MLV system, the elastic strain was measured directly by performing recovery experiments rather than estimated separately, providing a particularly stringent test of this criterion. The threshold stress separating the two branches, first identified by Coussot and coworkers [41]. and termed viscosity bifurcation, itself grows with the time the sample is allowed to age before the stress is applied. Its aging branch, on the other hand, on which no steady state is ever reached, is direct evidence for Definition 2 rather than for Definition 1 itself. Under Definition 2, various systems, including the aging branches already noted above, were shown to possess a creep compliance or relaxation modulus that continue to depend explicitly on waiting time without any sign of saturation. In most cases, these experiments led to a power-law growth of the material's underlying relaxation time, becoming unbounded as the waiting time is increased without bound.

Finally, we believe that strongly thixotropic materials will likely satisfy both definitions 1 and 2, but in ambiguous cases where distinguishing thixotropy from viscoelasticity becomes more difficult, we might find that one of our definitions is satisfied while the other is not. We see this layered structure as an advantage rather than a limitation of this approach. We believe it enables the use of the term 'thixotropic' with varying degrees of confidence, from materials that clearly meet both definitions to borderline cases that satisfy only one. Going forward, these definitions can be challenged, improved upon, and used, not only to classify a material as thixotropic but also to specify the particular meaning of the term being referred to.

## IV. Concluding Remarks

A hundred years is a long time for a technical term to drift without anchor. Building on Schalek and Szegvar's [18] report that an iron-oxide sol solidified at rest and reversibly liquefied on shaking, Péterfi [3] named this phenomenon 'thixotropy.' Over the century since, the concept of "thixotropy" has been broadened, in some cases to encompass almost any time-dependent change in viscosity, including ordinary nonlinear viscoelastic behavior. In Section I, we provide a historical account of how definitions of thixotropy have evolved over the past century. The purpose of this manuscript is not to supply definitions that put an end to discussion, but to guide the community closer to consensus through review of the historical record and the practical demands of simplicity, quantifiability, and agnosticism about microstructure, in accord with general rheological taxonomy. The principal message of this work is that criteria for thixotropy should be more definitive than seems to be the case at present, in light of the observation that time-dependent rheological responses can also arise from nonlinear viscoelasticity. Thus, a clearer distinction between these sources of nonlinear time dependence in rheology seems to be needed. The decisive distinction we attempt to draw here is whether the observed time dependence exceeds that attributable to viscoelasticity and reflects an evolving structural memory. Here "structure" is taken to be distinct from polymer-like orientation that gives rise to large recoverable strains and gradual stress relaxation after cessation of flow. Accordingly, in Section II, we have set out three definitions that make explicit the different (so far largely implicit) senses in which the term has actually been used. The first of these, Definition 0, is an idealized, purely inelastic limit, which serves mainly as a conceptual anchor for constitutive modeling. The remaining two definitions, on the other hand, translate directly into laboratory protocols and were tested, in Section III, against real materials. We note, however, that materials meeting Definition 1 contain stress responses to straining that are not accompanied by significant changes in recoverable elastic strain. These stresses can be modelled, at least approximately, by a purely viscous, time-dependent constitutive model that satisfies Definition 0. Thus, there is a practical connection between Definitions 0 and 1.

Definition 1 proposes that in a thixotropic material, following a sufficiently large increase in deformation rate or applied stress, the strain required to attain a steady-state greatly exceeds that associated with nonlinear viscoelasticity. Thus, according to Definition 1, there is a need to establish as a reference state the strain scale associated with reversible elastic deformation. The relevant comparison is, therefore, between the strain required for the material to reach steady state and the strain that its own viscoelasticity can account for. This distinction is essential because recoverable or elastic strain can vary substantially in magnitude across different materials. A magnitude of ten strain units may greatly exceed the elastic strain in a particulate suspension but be comparable to that in a highly entangled polymer. We therefore encourage researchers to measure a material-specific recoverable strain, or an equivalent elastic-to-viscous crossover scale, whenever possible, and to report it alongside any claim of thixotropy, especially if there is significant doubt regarding the magnitude of that recoverable strain for the type of material tested. Moreover, non-monotonicity of the steady-state flow curve and the corresponding occurrence of shear banding should not by itself be regarded as proof of thixotropy. Non-monotonicity of the flow curve could well be a consequence of physical aging–mechanical rejuvenation dynamics but is not itself a defining criterion of them. Furthermore, while selecting shear rates for the step jump, one must carefully select the shear rate range that falls within the monotonic regions of the flow curve. This ensures that the measured strain reflects genuine microstructural kinetics rather than the migration of interfaces or band-width readjustments.

Under stress control, Definition 1 recovers viscosity bifurcation as an important special case. Below a critical stress, physical aging leads to an unbounded growth of viscosity and relaxation time, while above it, the material reaches a steady flowing state, but only after strains far exceeding those attributable to nonlinear viscoelasticity. On its flowing branch, this stress-controlled form employs the same strain criterion as the rate-controlled form Definition 1, now applied to changes in stress rather than in shear rate. In this framework, the stress at which bifurcation between eventual fluid and solid behavior occurs is expressed as the critical stress. This critical stress differs from a conventional plastic yield stress below which material undergoes only the elastic deformation. It is, therefore, recommended that researchers report the entire experimental protocol, including the nature of any preshear, the elapsed waiting times, the applied stresses, and the corresponding duration of observation that differentiates the transient response, either a true arrest or a flowing state. It should be kept in mind that in the vicinity of an apparent critical stress, finite experimental windows can be profoundly misleading. This is because a material can, in principle, appear to flow for a substantial period before ultimately becoming arrested. We note, finally, that the aging branch of this bifurcation, on which no steady state is ever reached, is not itself evidence for Definition 1 in the strict sense; it is, instead, direct evidence for Definition 2.

According to Definition 2, thixotropy is exhibited when, after cessation of preshear, the material's relaxation time grows without bound during rest, provided this growth can be at least partially reversed by resumption of flow. Accordingly, the waiting-time dependence is in the relaxation timescale itself, rather than inferred from the growth of a storage modulus at any frequency or yield stress measured at different arbitrary probe times. This is because a plateau in $G'$, $G''$, or yield stress for a certain experimental condition is not sufficient evidence that structural evolution has ceased. Accordingly, the most convincing experiments measure a response function, rheological or otherwise, and its time dependence as a function of elapsed time over a broad range of waiting times. This facilitates extraction of the waiting time dependence of the characteristic relaxation time. Where this is not possible, the limitation should be stated explicitly, and such measurements be taken as possible *diagnostics* of thixotropy, rather than directly satisfying a definition of it. According to Definition 2, evidence for absence of saturation in relaxation time over prolonged waiting times is an essential criterion to meet the definition of thixotropy. We freely grant, however, that experimental evidence may accumulate that allows one to reliably infer thixotropy from behavior that can be shown to strongly correlate with behavior that meets this definition of thixotropy. We believe that distinguishing between definitions and diagnostics of "thixotropy" will be important in clarifying which rheological phenomena are common in "thixotropic" materials, but might be observed in purely viscoelastic ones as well.

We note that Definition 2 covers the rheological behavior of both hard and soft "glassy" materials. Workers in those fields are likely to refer to behavior satisfying our Definition 2 as "glassy" rheological behavior, especially when seen in conventional glasses, and even in pastes and emulsions modeled by what is called "Soft Glassy Rheology (SGR). However, when this behavior is observed in suspensions of hard particles, cellulose nanocrystals, clays, latex particles, and microgels exemplified here, it has frequently been referred to as "thixotropic," and so we deploy Definition 2 to cover these situations, and because many materials that satisfy Definition 1 and likely to satisfy Definition 2 as well.

Notwithstanding which definitions are being experimentally validated, while dealing with potentially thixotropic materials the initial state of the sample is part of the experiment and must be specified explicitly. A thixotropic response is intrinsically history dependent. Consequently, the phrase 'the material at a waiting time $t_w$' has little meaning unless the state from which $t_w$ is measured is clearly defined. Therefore, many details, such as preshear magnitude and duration, geometry, temperature, loading history, sample age, and the criterion used to decide that the material has been rejuvenated, matter when assessing thixotropic behavior of a material. Therefore, the protocol used is not just important for experimental housekeeping but as a part of the rheological specification of the material. We would like to emphasize that the reproducibility in measurements of thixotropy begins with reproducibility of the initial state. This is especially important when comparing experimental behavior of the same material from different laboratories. Owing to their being thermodynamically out of equilibrium and therefore strongly history dependent, nominally identical thixotropic samples can otherwise correspond to different points on a structural/energy landscape trajectory [96].

Although intended to be agnostic with respect to the underlying microstructural changes, thixotropy is not applicable to rheological time dependence caused by chemical reactions. In part, this is because chemical changes are irreversible, or at least not reversible by flow. Rheological time-dependence induced by temperature-induced phase change is also not considered "thixotropic." However, once the phase change has occurred, it might produce a thixotropic material. An example is temperature-induced crystallization, such as in waxy oils, which arise in pipeline flows [8, 9]. Once one quenches a waxy oil from a high temperature at which all wax molecules are dissolved to a lower one wherein wax crystals begin to grow, the formation and breakage of aggregates and networks of crystals becomes sensitive to flow history. This sensitivity, at constant temperature, produces long-lived rheological transients that satisfy one or more of the definitions of thixotropy given above. In particular, the changes in rheology brought about by crystal aggregation is to a large extent reversible by imposition of a strong flow, even at fixed temperature. Thus, although the initial formation of the dispersion of wax crystals in a solvent medium is induced by a temperature change, the phenomenology and even the microstructural changes brought about by changes in shear rate at fixed temperature are analogous to those observed in suspensions of aggregating particles. If the time-dependent changes in rheology are induced by both flow and temperature changes, which is relevant to flow of initially hot oil through cold pipelines, this behavior is referred to as "thermokinematic memory," [9] the thixotropic component of which is contained in the response to mechanical deformations under isothermal conditions.

Another case requiring some discussion is that of two-phase liquid-liquid suspensions, produced, for example by quenching a miscible multicomponent mixture into a two-phase region. After the quench, droplets or domains of one phase in another grow with time, and their microstructure and rheology becomes sensitive to shear rate and time. However, in this case, unless there is some means by which the growth of droplets or domains is arrested at long times under quiescent conditions, the system will eventually develop macroscopic regions that separate from each other under gravity, eventually forming two separate bulk phases. This situation is not one that is well described as "thixotropic." Thixotropic materials, of course, do include two-phase systems, especially in which one phase consists of solid particles. But even here, if the particles are large enough and dense enough to sediment, then the changes in macroscopic rheology that occur during sedimentation in a rheometer would be rightly excluded from the definition of "thixotropy." A "thixotropic" material must be one in which a dispersed phase is confined to domains small enough that sedimentation is avoided, so that material remains homogenous on scales much larger than the domain size. In practice, this requires domain sizes of no more than a few tens of microns. As an example of such a case, stabilized emulsions, such as mayonnaise, contain two liquids, but surface-active agents keep the structure from coarsening beyond scales of around 1-10 microns, and so they can qualify as "thixotropic" by virtue of their sensitivity to flow history. All thixotropic materials display time-dependent rheology that is substantially reversible, with a high viscosity or yield stress developing at rest or under weak shear rate or stress, and a lower viscosity under high shear rate or stress, and the ability to go back and forth between by changing flow conditions alone. Thus, the defining feature of thixotropy is not just that a material changes with time, but that the relevant rheological evolution is at substantially reversible under cycles of low (or zero) and high rates of deformation.

It is also important to comment on the relationship between the two operational definitions (1 and 2). For materials that show a viscosity bifurcation, the two definitions are closely connected. The stress controlled form of Definition 1 governs the flowing branch, on which a steady state is reached only after an anomalously large strain. Definition 2, on the contrary, governs the aging branch, on which viscosity and relaxation time grow without bound and no steady state is ever reached. In this sense, bifurcating materials are complementary manifestations of the same underlying competition between structural rebuilding and deformation-induced breakdown, probed on either side of a shared threshold. The rate-controlled form of Definition 1, however, does not require this connection. According to it a material requires an anomalously large strain to reach a new steady state after a step-up in shear rate. In principle, a genuinely, strongly thixotropic material should satisfy both definitions. However, the fact that a material may satisfy one definition more directly than the other, particularly where a rate-controlled test of Definition 1 has been performed without a corresponding stress-controlled or waiting-time-resolved study, should not necessarily be regarded as a contradiction. Instead, it may simply reflect which part of the material's behavior has been probed. Establishing when satisfying one definition implies the other, and identifying classes of materials for which no such implication holds, would be an interesting exercise to assess the validity, and the limits, of this framework.

We now raise a more philosophical discussion issue: what is evolving when a thixotropic material ages? The present framework deliberately avoids answering this question because doing so would make the definition dependent on microstructure. As mentioned before, the thixotropic materials are thermodynamically out of equilibrium soft materials that do not reach the equilibrium state over practical timescales. Consequently, the answer to the above question certainly is that the free energy of a thixotropic material progressively decreases under quiescent conditions as time passes. However, in rheological language, the continued growth of relaxation times, the evolution of elastic modulus and yield stress, and/or the passage through a critical gel state in addition to the persistence of post-gel dynamics, all suggest that a boundless evolving spectrum of relaxation modes may be intrinsic to thixotropic character of a material.

Finally, the framework proposed in this work should not be regarded as a final taxonomy on thixotropy. There is a possibility to encounter systems for which the proposed three definitions are not sufficient. Such failures need not, in principle, invalidate the framework, nor would they undermine the possible thixotropic character of a material. What matters is that a new classification should make explicit what rheological measurements and observations warrant the application of the term

“thixotropy” rather than some other rheological term. This requires that a reference response be defined that must clearly be met in a material, and there is no better rheological descriptor than “thixotropy” already available to describe it. This could involve a large enough timescale or strain scale that justifies excluding the alternative rheological descriptor, usually viscoelasticity, or in some cases, plasticity. Extensive modeling efforts have sought to combine thixotropic and viscoelastic descriptors within a single framework [11, 15, 25, 97-101]. Ewoldt and McKinley [102], instead proposed mapping a material's position within a continuous, dimensionless thixo-elasto-visco-plastic parameter space, while Sen and Ewoldt [103] developed a model-insensitive framework, based on thixotropic spectra extracted from step-shear data, for comparing thixotropic properties across materials on Ashby-style charts. Looking ahead we hope that this discipline will make it possible to reserve the term "thixotropy" for a family of experimentally distinguishable rheological behaviors similar to the examples discussed in this work. We emphasize that an enduring value of the concept of thixotropy need not lie in finding a single universal definition. Rather one could establish multiple phenomenological expressions for thixotropy, as done in this work, by which rheologists can distinguish thixotropy from other forms of time-dependent rheology. Whether there really are different types of thixotropy, or that all definitions proposed here overlap so thoroughly as to make them functionally equivalent, remains to be determined. In that sense, the objective of this work is not to close a long-standing debate, but to provide a direction for future, more rigorous, work.

**Acknowledgments:**

YMJ acknowledges financial support from the Anusandhan National Research Foundation, Government of India (under grant JCB/2022/000040). He also acknowledges correspondence with Prof. Philippe Coussot and Prof. Daniel Bonn. RGL acknowledges funding from the National Science Foundation under grant CBET-2323147. Any opinions, findings, and conclusions or recommendations expressed in this material are those of the authors and do not necessarily reflect the views of NSF.